\documentclass[preprint,12pt,authoryear]{elsarticle}
\usepackage{graphics,bm,graphicx}
\usepackage{float}
\usepackage{epstopdf}
\usepackage{tikz}
\usepackage{pdfpages}

\usepackage{subfig}
\usepackage{mathtools}
\usepackage{amsmath,mathptmx}
\usepackage[displaymath,mathlines]{lineno}
\usepackage{amssymb}
\usepackage{multicol}
\usepackage{booktabs}

\usepackage{timet}
\begin{document}

\begin{frontmatter}

\title{The role of slow magnetostrophic waves in the formation of 
	the axial dipole in planetary dynamos}
\author{Aditya Varma}
\author{Binod Sreenivasan}

\ead{bsreeni@iisc.ac.in}



\address{Centre for Earth Sciences, 
Indian Institute of Science, Bangalore 560012, India}
	%
\begin{abstract}
The preference for the axial dipole in planetary dynamos is
investigated through the 
analysis of wave motions in spherical dynamo models. Our
study focuses on the role of slow magnetostrophic waves,
which are generated from localized balances between the
Lorentz, Coriolis and buoyancy (MAC) forces.
Since the slow  
	waves are known to 
	intensify with increasing field
	strength, simulations in which the field grows
	from a small seed towards saturation are useful
	in understanding the role of these waves in dynamo
	action.  Axial
	group velocity measurements in the energy-containing
	scales show that 
	fast inertial waves slightly modified by the magnetic field 
	and buoyancy are dominant under weak fields. However,
	the dominance of the slow waves is evident for strong fields
	satisfying $|\omega_M/\omega_C| \sim $ 0.1, where $\omega_M$ and
	$\omega_C$ are the frequencies of 
	the Alfv\'en and inertial waves respectively. A MAC wave window
	of azimuthal wavenumbers is identified wherein helicity
	generation by the slow waves
	strongly correlates with dipole generation. 
	Analysis of the magnetic 
	induction equation suggests a poloidal--poloidal
	field conversion in the formation of the dipole. Finally, 
	the attenuation of
	slow waves may result in 
	polarity reversals in a strongly driven Earth's core. 

\end{abstract}

%
\end{frontmatter}	
	
	\section{Introduction}
	
	Planetary dynamos are driven by thermochemical convection
	in their fluid cores. The axial dipole dominates
	a large region of the parameter space in convection-driven
	dynamos where the effect of planetary rotation, measured
	by the Coriolis forces, is large relative to that of 
	both nonlinear inertia and viscosity \citep{
	06sreejon,11sreeni,schaeff2017}. Rapid rotation produces
	anisotropic convection with equatorially antisymmetric axial
	motions, the helicity of which is thought to be essential
	for dynamo action \citep{moffatt1978, 
	olson1999numerical}. A long-standing question in planetary dynamo
	theory is whether the preference for the
	axial dipole is due
	to a purely kinematic process influenced by rotation
	or due to a magnetohydrodynamic process influenced by both rotation
	and the self-generated magnetic field. Answering this question
	would also help us constrain the parameter space that
	admits polarity reversals in strongly driven
	dynamos \citep[e.g.][]{14sreeni}.

	An early study by \cite{76busse} used the linear
	theory of magnetoconvection to explore the onset of
	dynamo action in an annulus. Busse found that the effect
	of a magnetic field on convection enhanced 
	magnetic field generation. This interesting idea was
	explored further by \cite{11sreeni} who showed that
	the presence of a magnetic field substantially enhanced
	the kinetic helicity of columnar convection. They considered
	linear magnetoconvection in a spherical shell in the
	rapidly rotating limit $E \to 0$, where $E$ is the Ekman number
	that gives the ratio of viscous to Coriolis forces.
	Although the spatially varying magnetic field in 
	a nonlinear dynamo 
	does not substantially lower the threshold for
	convective onset relative to that in the nonmagnetic
	system \citep{jfm17a}, there is a substantial
	enhancement of helical convection 
	in the neighbourhood of the length scale of energy injection
	\citep{prf2018}. The growth of convection is
	notably absent in a kinematic dynamo, which fails
	to produce the axial dipole with the same parameters
	and initial conditions. While nonlinear dynamo models
	strongly relate field-induced helicity generation 
	in the energy-containing scales 
	to dipole formation, the primary force
	balance in these scales is known to be approximately
	geostrophic \citep{aurnou2017,aubert2017}, which raises the
	question of how the field acts on these scales so as to
	enhance helicity. The present study addresses this
	question by analyzing wave motions in the energy-containing
	scales in planetary dynamo models.  
	
	Wave motions in planetary cores arise from the 
	effects of rotation, magnetic field and buoyancy.
	Torsional oscillations propagating radially at the
	Alfv\'en speed 
	across concentric
	cylinders have been simulated in low-inertia numerical
	models of the geodynamo
	 \citep{wicht2010torsional, teed2014dynamics}.
	Non-axisymmetric Alfv\'en waves propagating along the cylindrical
	radius are conceivable \citep{08jault, 16bardsley,
	 aubert2019} since the convection
	 is made up of thin columns aligned with the rotation
	 axis. Slow Magneto-Coriolis (MC)
	   Rossby waves, thought to produce the westward drift of the Earth's
	   magnetic field, have been realized in dynamo
	   simulations 
	\citep{hori2015slow}. While convection can onset
	in the form of Alfv\'en waves in a non-rotating
	B\'enard layer \citep{00robzhang},
	the planetary regime of strong 
	rotation can support convection through fast and slow
	Magnetic-Archimedean-Coriolis (MAC) waves. The
	fast MAC waves are inertial waves weakly modified
	by the magnetic field and buoyancy; the
	slow MAC, or magnetostrophic, waves
	 are slow MC waves modified
	by buoyancy \citep{67brag,07bussechapter}.
	Buoyancy-driven fast inertial
	waves generate and segregate oppositely signed
	helicity in spherical dynamos \citep{ranjan2018internally}.
	That said, the intensity of slow MAC wave motions
	would be comparable to that of the fast waves for
	$|\omega_M/\omega_C| \sim$ 0.1, where $\omega_M$ and
	$\omega_C$ are the Alfv\'en wave and inertial wave
	frequencies respectively \citep{jfm21}. Here, we examine
	the role of the slow MAC waves in helicity generation in
	the energy-containing scales of the dynamo, and
	hence in axial dipole formation. While  earlier studies
	have related axisymmetric MAC waves in the 
	stably stratified layer at the top of the core to the decadal 
	oscillations in the Earth's field \citep{buffett2016},
	the focus of the present study is to relate non-axisymmetric
	MAC waves
	in an unstably stratified core to the formation of
	the dipole field. Because slow MAC
	waves intensify with increasing field
	strength, a nonlinear simulation in which the field grows
	from a small seed towards saturation would help us understand
	when the slow waves have a dominant presence alongside 
	the fast waves in the
	dynamo.
	 
	Numerical dynamo models \citep{olson1999numerical, 
		kageyama1997generation} have shown how columnar vortices
		twist the toroidal magnetic field
	lines to produce the poloidal field.
	\cite{takahashi2012detailed} and \cite{Pena2018Dec} made a
    detailed analysis of terms in the magnetic
    induction equation. The present study looks at
    the dominant contributions to the axial dipole field and brings
    out the differences between
	kinematic and nonlinear dynamos in this respect.
	
	In Section 2, we describe the dynamo model and define
	the main dimensionless parameters used in this study. 
	Section 3 builds on a recent study that suggested
	field-induced 
	helicity generation in the relatively large scales of
	the dynamo \citep{prf2018} and
	shows through force balances that local magnetostrophy
	can exist in these scales where the Lorentz forces are
	small in the global balance. Section 4 analyses the
	fundamental frequencies in the dynamo and shows that the
	MAC wave window of azimuthal wavenumbers is indeed where
	the axial dipole is predominantly generated. In Section 5,
	the slow MAC waves in nonlinear dynamo simulations
	are identified by group velocity measurements. Section 6
	gives the contributions to the axial dipole
	of the dominant terms in the
	induction equation in nonlinear and kinematic dynamo
	simulations. In conclusion, 
	the main results of this study are summarized and its implications
	for polarity reversals in strongly driven dynamos are discussed.
	
	\section{Numerical dynamo model}
	
	We consider dynamo action in 
	an electrically conducting fluid confined
	between two concentric, corotating 
	spherical surfaces that correspond to the inner core
boundary (ICB) and the CMB. The ratio of inner to outer radius is
0.35. Fluid motion is driven by 
thermal buoyancy-driven convection, although our set of equations
may also be used to study thermochemical convection using
the codensity formulation \citep{95brag}. The other
body forces acting on the fluid are the Lorentz force, arising from
the interaction between the induced electric currents and the magnetic
fields and the Coriolis force originating from the background
rotation. The governing equations are those in 
the Boussinesq approximation \citep{02konorob}. 
Lengths are scaled by the thickness of the spherical shell $L$, and time is scaled 
by the magnetic diffusion time, $L^2/\eta$, where $\eta$ is the magnetic diffusivity.  
The velocity field ${\bm u}$ is scaled by $\eta/L$, 
the magnetic field ${\bm B}$ is scaled by 
$(2\varOmega\rho\mu\eta)^{1/2}$ where $\varOmega$ is the rotation rate, 
$\rho$ is the fluid density and $\mu$ is 
the magnetic permeability. 
The root mean square (rms) and peak
values of the scaled magnetic field (Elsasser number $\varLambda$)
are outputs derived from  our dynamo simulations, 
where the mean is a volume average. 	

The non-dimensional MHD equations for the velocity, 
magnetic field and temperature are given by,	

	\begin{align}
E Pm^{-1}  \Bigl(\frac{\partial {\bm u}}{\partial t} + 
(\nabla \times {\bm u}) \times {\bm u}
\Bigr)+  {\hat{\bm{z}}} \times {\bm u} = - \nabla p^\star +
Ra \, E \, Pm Pr^{-1} \, T \, {\bm r} \,  \nonumber\\ +  (\nabla \times {\bm B})
\times {\bm B} + E\nabla^2 {\bm u}, \label{momentum} \\
\frac{\partial {\bm B}}{\partial t} = \nabla \times ({\bm u} \times {\bm B}) 
+ \nabla^2 {\bm B},  \label{induction}\\
\frac{\partial T}{\partial t} +({\bm u} \cdot \nabla) T =  Pm Pr^{-1} \,
\nabla^2 T,  \label{heat1}\\
\nabla \cdot {\bm u}  =  \nabla \cdot {\bm B} = 0,  \label{div}
\end{align}
	
The modified pressure $p^*$  in equation \eqref{momentum}
 is given by $p+E \, Pm^{-1} \, |\bm{u}|^2$.
The dimensionless parameters in the above equations 
are the Ekman number $E=\nu/2\varOmega L^2$,
the Prandtl number, $Pr=\nu/\kappa$, 
the magnetic Prandtl number, $Pm=\nu/\eta$ and
the Rayleigh number $g \alpha L\Delta T/2 \Omega \kappa$. 
Here, $g$ is the gravitational acceleration, 
$\nu$ is the kinematic viscosity, 
$\kappa$ is the thermal diffusivity and $\alpha$ 
is the thermal expansion coefficient.

The
basic-state temperature profile represents a basal heating
given by $T_0(r) = \beta/r$, where $\beta$ is a constant. We set
 isothermal conditions at both boundaries. The velocity and 
magnetic fields satisfy the no-slip and electrically
insulating conditions, respectively \citep{06sreejon}.
The calculations are performed by a pseudospectral 
code that uses
spherical harmonic expansions in the angular coordinates $(\theta,\phi)$
and finite differences in radius $r$ \citep{07willis}.

\section{Helicity generation during magnetic field growth from a seed}

\begin{table*}
\caption{Summary of the key important input and output
parameters in the dynamo simulations considered in the present study.
Here, $N_r$ is the number of radial grid points, $l_{max}$
is the maximum spherical harmonic      degree, 
$\bar B$ is the volume-averaged root mean square value
of the magnetic field, $B^P_{10}$ is the poloidal 
axial dipole field, $Ra_c$ is the critical Rayleigh number for
the onset of nonmagnetic convection, $Rm$ is the magnetic Reynolds number, and
$l_{c}$ and $l_{E}$ are the mean spherical harmonic degrees
for convection and energy injection respectively, defined
in \eqref{elldef}. The values in brackets 
are those for the reference nonmagnetic runs.}

\begin{minipage}{1\linewidth }
	\centering
	\resizebox{\columnwidth}{!}{
	\begin{tabular}{c c c c c c c c c c c c c}
		
		S.No. & $E$ & $Ra$ & $Ra/Ra_c$ & $Pm$ & $Pr$ 
		& $N_{r}$ & $l_{max}$& $\bar B$ & 
		$B^P_{10}$& $Rm$ & $l_{c}$ & $l_{E}$ \\
		\hline
		
		a  & $1.2 \times 10^{-5}$ & 220 & 4.2& 5 & 5 & 120 & 100 & 0.7 & 0.30& 105   & 21 (28) & 20 (29)  \\
		b& $1.2 \times 10^{-5}$ & 500 &9.6  & 5 & 5 & 144 & 120 & 1.68 &0.80& 184  & 23 (34) & 23 (32) \\
		c & $1.2 \times 10^{-5}$ & 1000 &19.2 & 5 & 5 & 168 & 176  & 2.31 & 0.88& 326 & 26 (35) & 29 (37) \\
		d  & $1.2 \times 10^{-5}$ & 2000 &38.5 & 5 & 5 & 192 & 224  & 3.05 & 0.96& 558 & 28 (39) & 33 (40) \\
		e  & $1.2 \times 10^{-5}$ & 5000 &96.1  &5 & 5 & 180 & 224  & 3.79 & 0.87& 1218  & 31 (22) & 42 (31)\\
		f  & $1.2 \times 10^{-5}$ & 15000 &288.4 & 5 & 5 & 288 & 280  & 6.21&  0.84& 2710 & 33 (21) & 46 (36) \\
		g & $1.2 \times 10^{-6}$ & 400 & 7.3 &1 & 1 & 192 & 220 & 0.86 & 0.3& 215 & 39 (43) & 31 (49) \\
		\hline
		
	\end{tabular}
}
\end{minipage}
\label{Tab:param}
\end{table*}
	In line with an earlier studies \citep{11sreeni,prf2018},
	we examine the evolution of the dynamo from an initial
	dipole-dominated seed magnetic field of
	intensity $\overline {B}=0.01$. 
	The initial velocity field is the
	 same as that in the equivalent
	saturated non-magnetic run.  
	The key input and output parameters
	of the simulations, given in table \ref{Tab:param},
	are time-averaged values in the saturated state of
	the dynamo. Here, the 
	mean spherical harmonic degrees for convection and
	energy injection are defined by
	\begin{equation}
	l_{c}= \dfrac{\Sigma \hspace{1pt} l \hspace{1pt}
	E_{k}(l)}  {\Sigma \hspace{1pt} E_{k}(l)}; \hspace{5pt}
	l_{E}= \dfrac{\Sigma \hspace{1pt} l \hspace{1pt}
	E_{T}(l)}{\Sigma \hspace{1pt} E_{T}(l)},
	\label{elldef}
	\end{equation}
	where $E_{k}(l)$ is the kinetic energy spectrum and
	$E_{T}(l)$ is the spectrum obtained from the product of $u_{r}T$
	and its conjugate.
	
\begin{figure}
	
	{
		\centering
		
		\subfloat[\label{fig:els_fdip_a}]{\includegraphics[width=0.45\linewidth]{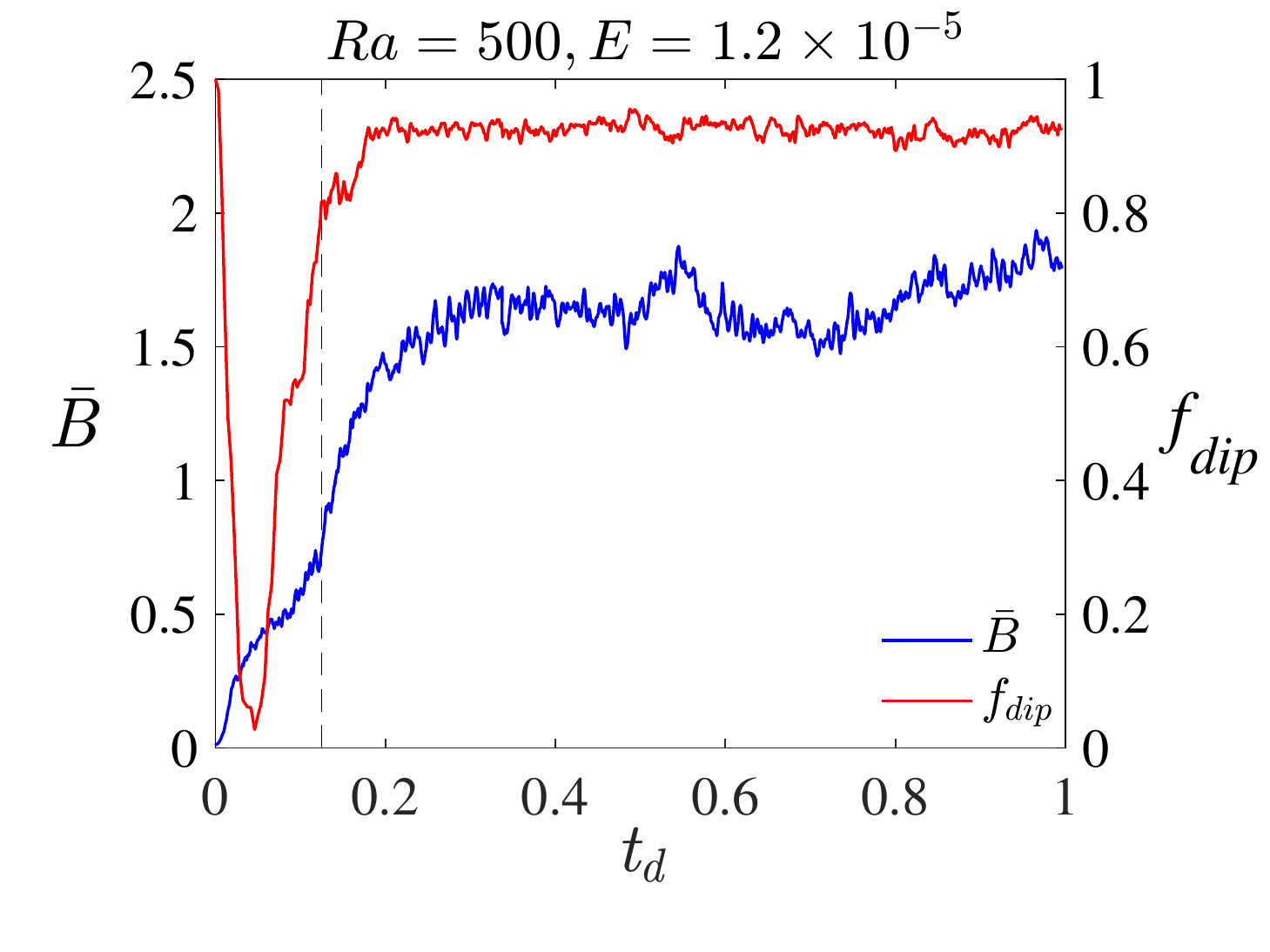}}		
		\subfloat[\label{fig:els_fdip_b} ]{\includegraphics[width=0.45\linewidth]{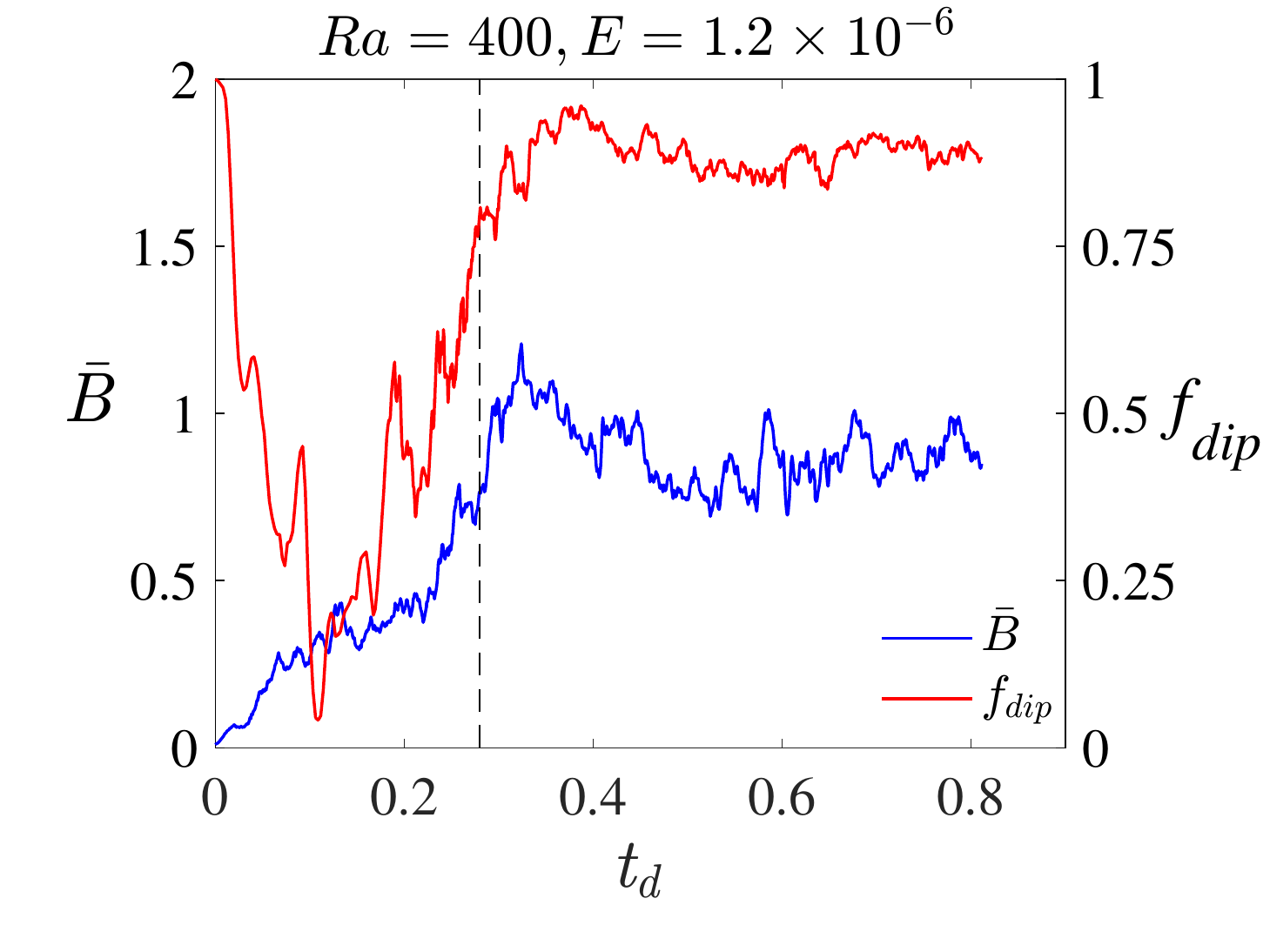}}
		
		\caption{Evolution in time (measured in units of
			magnetic diffusion time) of the magnetic field
			intensity given by its volume averaged
			root mean square value, $\bar B$ 
			and $f_{dip}$ (a measure of the axial dipole strength).
			The dipole formation time, marked by the vertical dashed
			line, is at $t_{d}=0.125$ in (a) and $t_{d}=0.28$ in (b). 
		}
		\label{fig:els_fdip}
	}
	
\end{figure}

\begin{figure}

{
	\centering
	\subfloat[ \label{fig:uz_rms_a} ]{\includegraphics[width=0.45\linewidth]{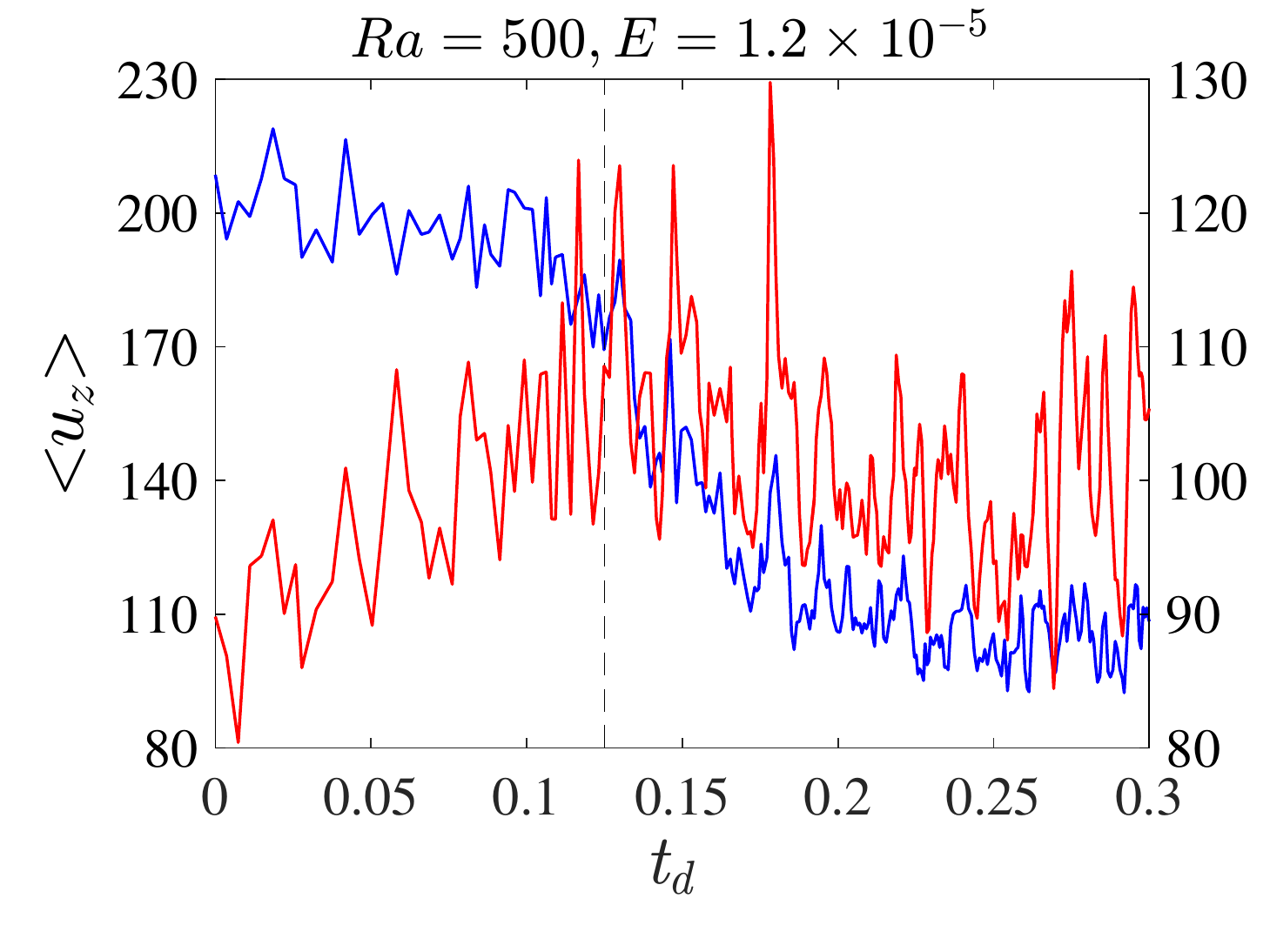}} 
	\hspace{1cm}
	\subfloat[ \label{fig:uz_rms_b}]{\includegraphics[width=0.45\linewidth]{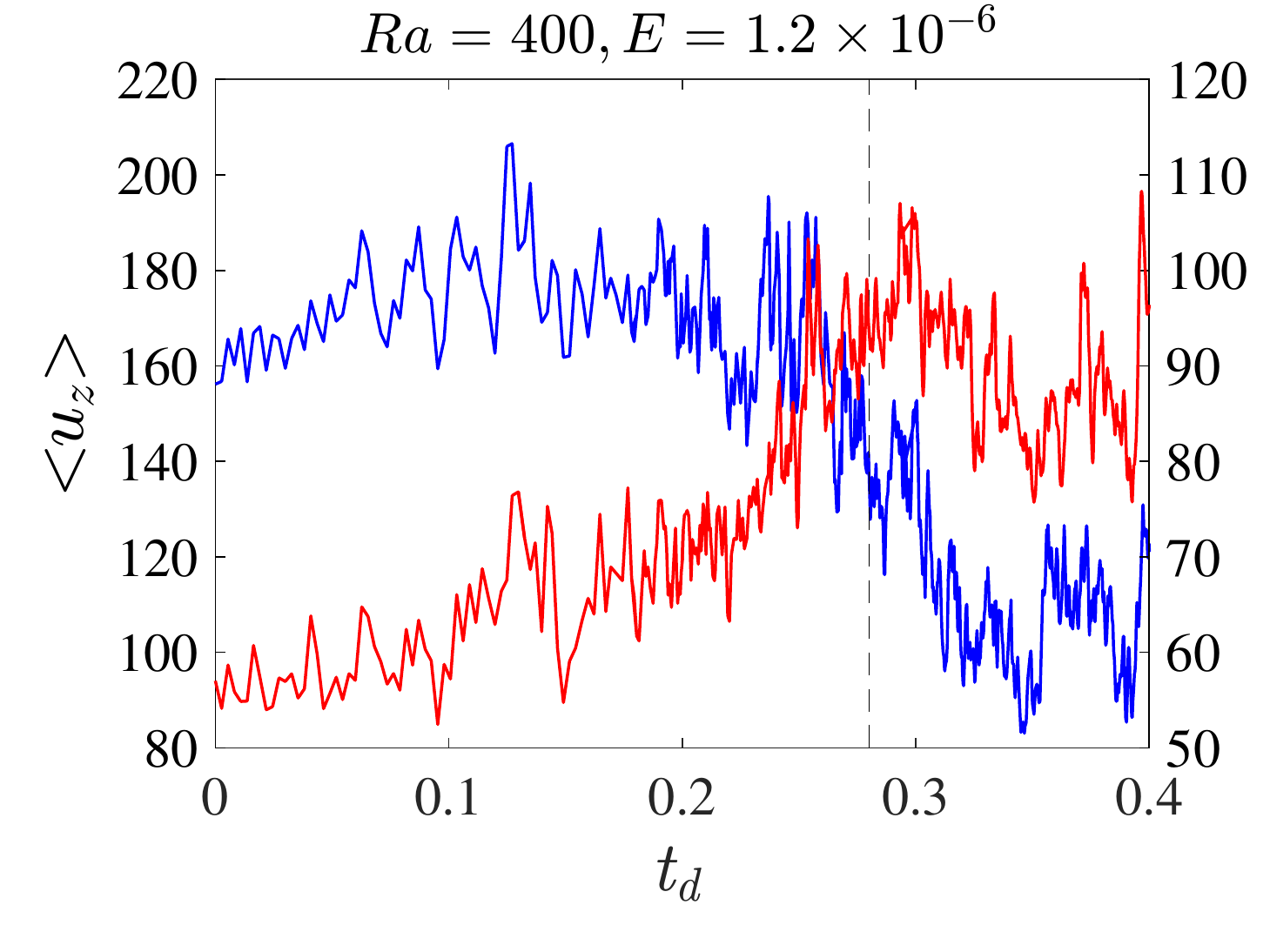}}\\
	
	\caption{Root mean square value of the axial
	velocity $u_{z}$ for two ranges
		of spherical harmonic degree, l. 
		The scales considered are $l \leq 23$ (red) in (a) and 
		 $l \leq 31$ (red) in (b),  $l > 23$ (blue) 
		in (a)  and $l > 31$ (blue) in (b). The mean harmonic
degree of energy injection $l_E$ serves as the basis for
		separation of scales.
		\label{fig:uz_rms}
	}
}

\end{figure}

	 The value of $f_{dip}$, which measures the relative energy contained in
the axial dipole \citep{06chraub}, 
	shows that the field loses its dipolar character and
	only regains it after passing through a non-dipolar phase 
	(figure \ref{fig:els_fdip}). The snapshots of the radial
	field at the core--mantle boundary during this process
	can be found in a previous paper \citep{prf2018} and hence
	not reproduced here.  Dynamo saturation occurs 
	only later than dipole formation. The time for formation of dipole 
	decreases at high $Ra$.  
	The progressive increase of the magnetic field intensity during
	dynamo evolution is
	accompanied by an increase of the convective
	velocity in the \lq\lq large" energy-containing 
	scales. The scales are separated
	by the mean harmonic degree of energy
	injection,  $l_{E}$. There is little or no
	increase of the velocity in the scales  $l> l_{E}$. 
	From figure \ref{fig:uz_rms}, we note that the 
	time of formation of the dipole
	roughly corresponds with the saturation	of the
	axial velocity $u_{z}$ 
	in the large scales. For the moderate forcing
	considered here ($Ra/Ra_c \sim 10$), the extraction of kinetic
	energy from the small scales, due to
	the Lorentz force occurs only near the formation 
	of the dipole. The magnetic field is fed by this kinetic
	energy but the growth of magnetic energy is not much 
	due to this process. The extraction of energy occurs
	in a relatively short time. Thus, the growth of energy in the large scales and 
	extraction of energy from the small scales remains fairly
	independent. We might hypothesize that a 
	quasi-linear wave excitation in the large scales
	of the dynamo would cause the enhancement of convective
	velocity over that in the equivalent nonmagnetic state.
	As the forcing is increased, the extraction of energy 
	from the small scales happens before the formation
	of the dipole. This would mean that at higher $Ra/Ra_c$, the growth of
	magnetic field is partially fed by the kinetic energy
	from the small scales. 
	
\begin{figure}
   {
		\centering	
	\subfloat[	\label{fig:hel_curvefit_a}]{\includegraphics[width=0.5\linewidth]{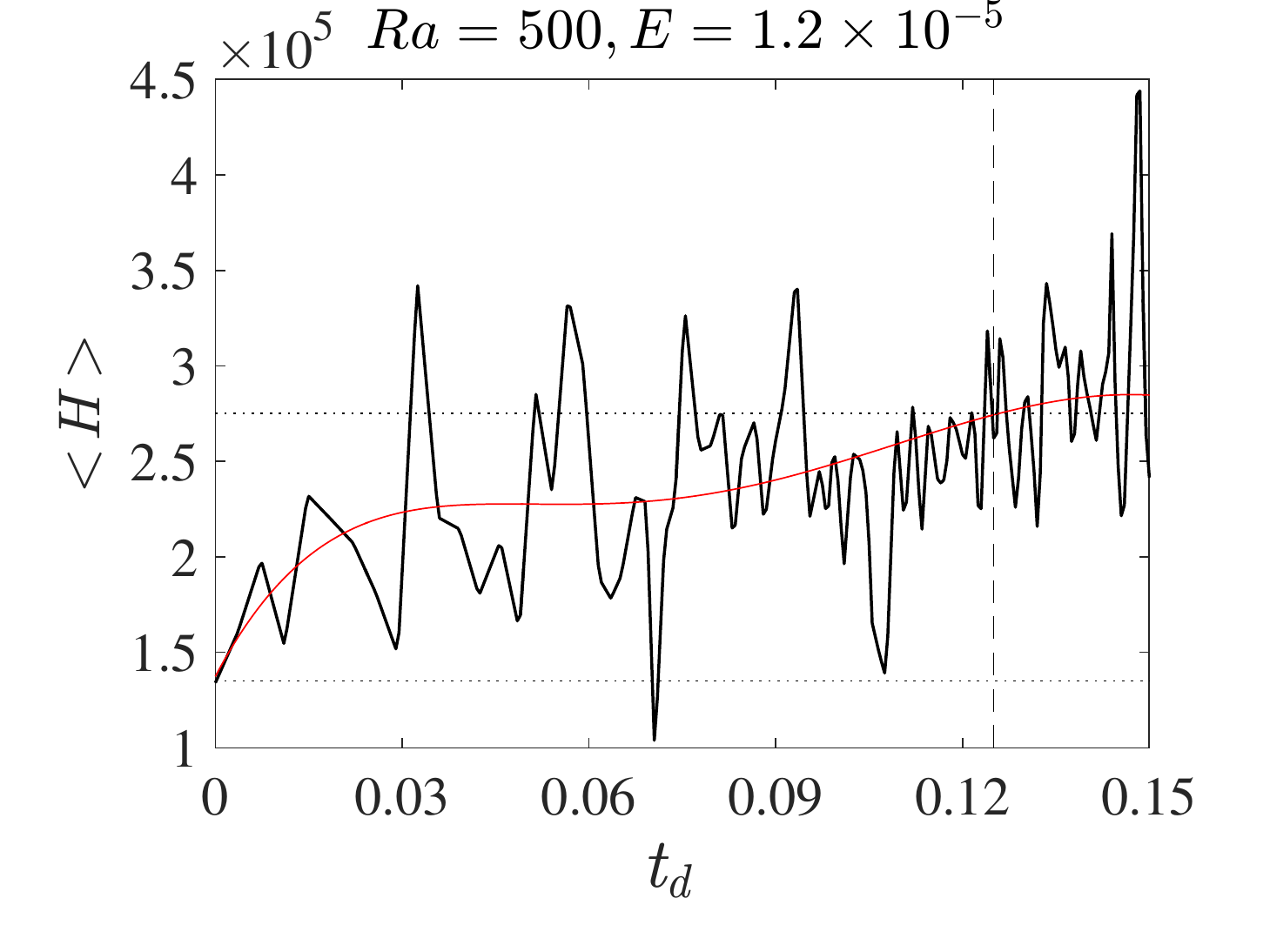}}	
	\subfloat[	\label{fig:hel_curvefit_b}]{\includegraphics[width=0.5\linewidth]{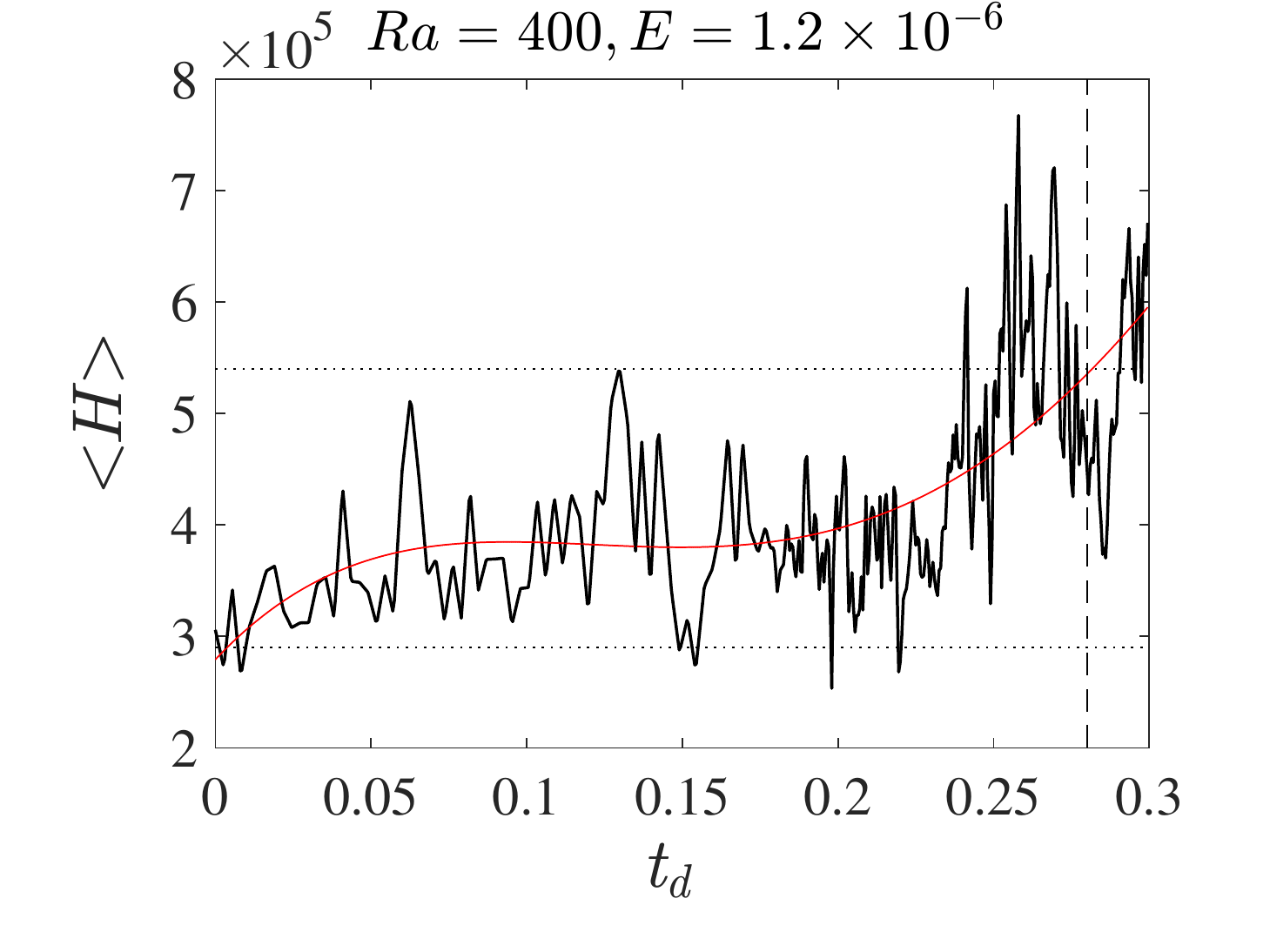}}\\
	\vspace{-0.2 in}
			\subfloat[	\label{fig:hel_l_c}]{\includegraphics[width=0.5\linewidth]{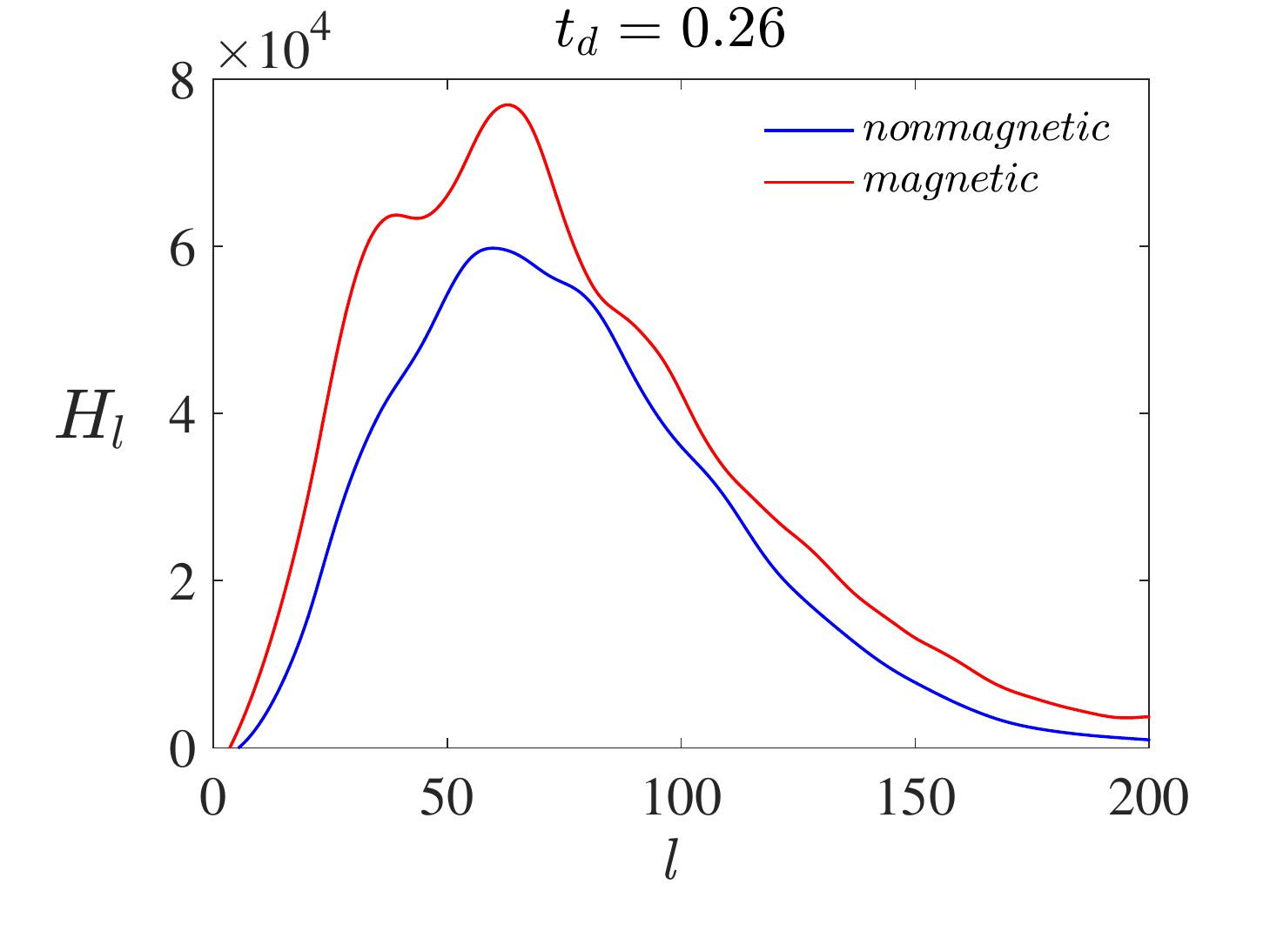}}	
		\subfloat[	\label{fig:hel_l_d}]{\includegraphics[,width=0.5\linewidth]{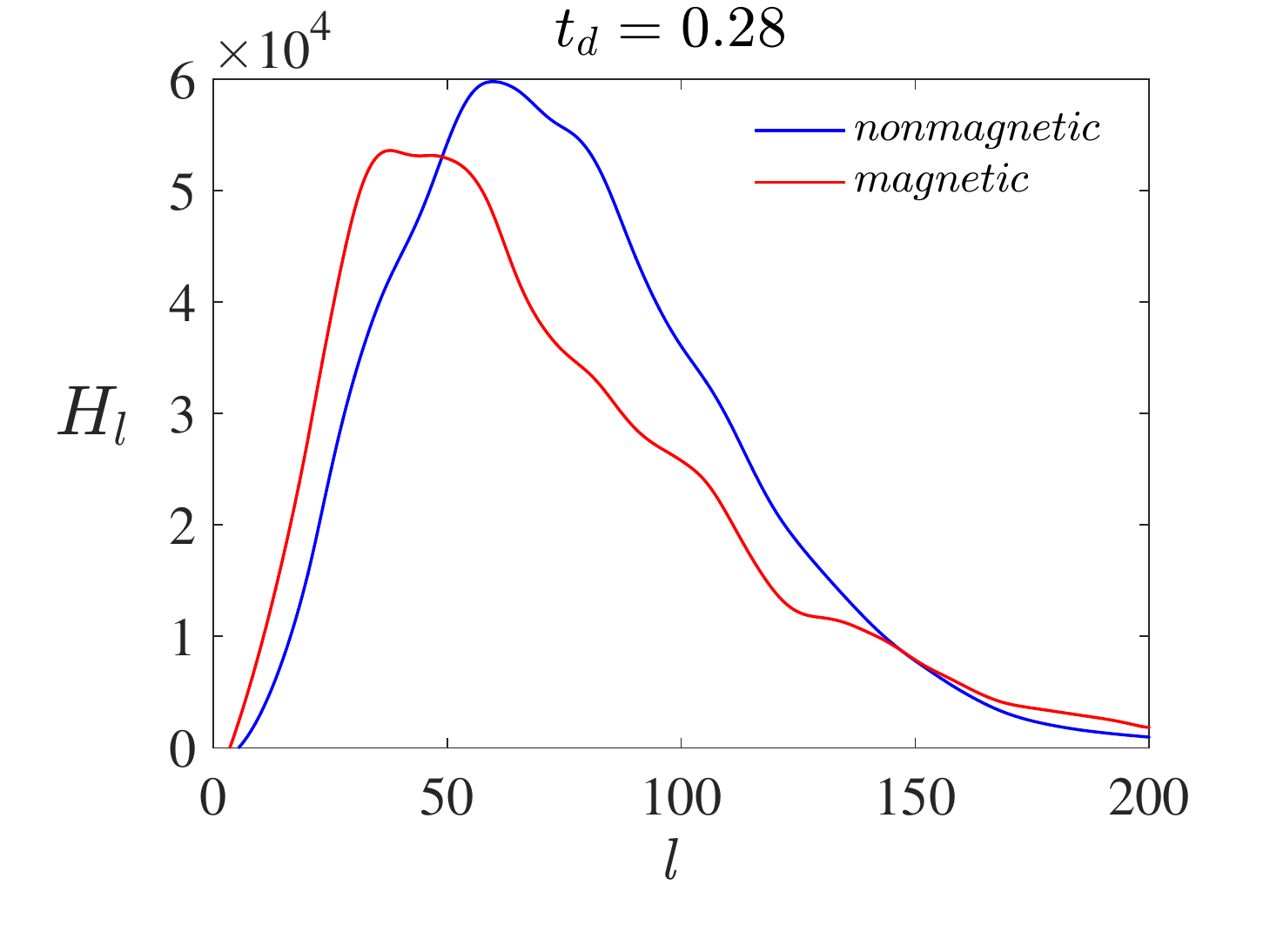}}

		\caption{(a) \& (b) Sum of the axial ($z$)
		 and radial ($s$) helicity $<\!\!H\!\!>$
			for the lower half of the spherical shell, plotted 
			against time (measured in  units of 
			the magnetic diffusion time $t_{d}$). The scales
			considered are $l \leq 23 $ for (a) and $l \leq 31$ for (b).
			The dynamo parameters are $Ra=500$, $Pm=Pr=5$,
			$E=1.2 \times 10^{-5}$ for (a) and $Ra=400$,
			$Pm=Pr=1$, $E=1.2 \times 10^{-6}$ for (b). The dashed
			vertical line indicates dipole formation time.
			(c) \& (d) Distribution of helicity over spherical
			harmonic degree at two times near dipole formation
			for the simulation in (b).}
		\label{fig:hel_curvefit}
	}	
\end{figure}

Figure \ref{fig:hel_curvefit} (a)  and (b) 
   show the enhancement of kinetic 
   helicity in the large scales for the two 
   dynamo simulations considered in figure \ref{fig:uz_rms}. 
   In cylindrical coordinates $(s,\phi,z)$,
    the magnetic field enhances the axial ($z$) and 
   radial ($s$) parts of the helicity in equal measure 
   \citep{11sreeni}. Therefore, the sum of the $z$ and $s$ 
   parts of the helicity is considered. 
    A notable finding is that the dipole
   forms from a chaotic multipolar state 
   when the helicity in the large scales increases by a
   magnitude nearly equal to the initial helicity 
   in these scales, associated with convection
    in the equivalent nonmagnetic
   system. Table \ref{Tab:z-hel} shows the  
   sum of peak $z$ and $s$ helicity attained during the growth of
   the magnetic field for the lower half of the shell. For
   moderate $Ra$, the total helicity over all
   scales during the growth of the magnetic
   field is higher than the nonmagnetic value 
   as extraction of kinetic energy from the small
   scales occurs only near the dipole formation time. 
   This is illustrated in figure \ref{fig:hel_curvefit} (c), where at 
   $t_{d}=0.26$, peak helicity growth occurs  such that 
   the helicity in the dynamo exceeds the nonmagnetic helicity
   for all scales. By time 
   $t_{d}=0.28$  (figure \ref{fig:hel_curvefit} (d)),
   energy extraction would begin and the helicity in the small
   scales would fall below the nonmagnetic values. 
   By the time the dynamo reaches saturation, the helicity in the small
   scales would have fallen even further.
   The helicity deficit in the small scales  dominates the
   helicity generated in the large scales and therefore
   the total helicity in the 
   saturated dynamo is always less than its nonmagnetic counterpart.
   For higher $Ra$, the total helicity would always be the less than
   that in the nonmagnetic case 
   as the energy extraction from the small scales occurs much before
   dipole formation. The helicity in the large scales
   would, however, still exceed the helicity of the nonmagnetic
   case. 	
   
   \cite{ranjan2020generation} show that the helicity source term
	due to the Lorentz force is negatively correlated with the 
	overall helicity distribution. They attributed the distribution
	of helicity in the core to inertial waves. No scale
	separation was performed, so the overall helicity 
	for the saturated dynamo was lower than that for
	the nonmagnetic state. Their result is consistent with our
	analysis considering all scales (table \ref{Tab:z-hel}).
	To show that the slow MAC waves 
	 might cause the increase of dynamo helicity
	over the nonmagnetic value at scales $l \leq l_E$, 
	one must first look at
	the scale-separated force balance in the dynamo.

\begin{table*}
	
	\caption{  Sum of the axial ($z$) and and radial ($s$) parts of the 
	kinetic helicity for the large (energy-containing)
		scales and for all scales evaluated at two
		times during the evolution of the dynamo magnetic field
		from a seed state.
		The helicity is evaluated for the lower half of the 
		spherical shell.
		The nonmagnetic helicity is given in
		brackets. The large scales are those for
		$l \leq l_E$, where $l_E$ is the mean harmonic degree
		of energy injection in the dynamo.}
	\centering
	\begin{tabular}{c c c c c c c c c c c c}
		
		S.No.  & $E$, $Ra$ & time ($t_{d}$) & Scales &  
		 Helicity \\
		\hline
		
		(i) & $1.2 \times 10^{-5}$ , 500 & 0.095 & $l \leq 23$ & $3.21 \times 10^{5}$ \hspace{0pt}$(1.34 \times 10^{5})$\\
		&  &  & All &$7.66 \times 10^{6}$  \hspace{1pt}$(7.23 \times 10^6)$\\
		 &  & 0.125    & $l \leq 23$ &$2.81 \times 10^{5}$ \hspace{0pt}$(1.34 \times 10^{5})$\\
		&  &  & All &$ 6.41\times 10^{6}$  \hspace{1pt}$(7.23 \times 10^6)$\\
	\hline
		(ii) & $1.2 \times 10^{-6}$, 400 & 0.26 & $ l \leq 31$ &$6.89 \times 10^{5}$ \hspace{0pt}$(3.05\times 10^5)$\\
		& &  & All &$7.15 \times 10^{6}$ \hspace{1pt}$(4.85\times 10^6)$\\
		 & &  0.28 & $ l \leq 31$  & $6.72 \times 10^{5}$ \hspace{0pt}$(3.05\times 10^5)$\\
		& &  & All &$3.89 \times 10^{6}$ \hspace{1pt}$(4.85\times 10^6)$\\
\hline		
	
	\end{tabular}
	\label{Tab:z-hel}
\end{table*}
\subsection{Scale-dependent balance of forces}


%

	Dipole-dominated dynamos are known to exist
	in the so-called MAC regime where 
	Lorentz, buoyancy and Coriolis forces are dominant \citep{06sreejon,11sreeni}
	and the nonlinear inertial and viscous forces are
	negligible. The Lorentz forces may, however,
	be localized due to spatially
	inhomogeneous magnetic flux. In line with earlier
	studies \citep{06sreejon}, we examine the ratio 
	of the Lorentz, Coriolis and 
	buoyancy forces in the $z$ vorticity equation
	to the highest force among 
	them for two distinct ranges of the spherical harmonic degree.

\begin{figure}	
{\centering
	
	\begin{minipage}{0.45\linewidth}
		\centering
		\subfloat[\hspace{1.5cm} M 	\label{fig:MAC_a} ]{\includegraphics[width=0.7\linewidth]{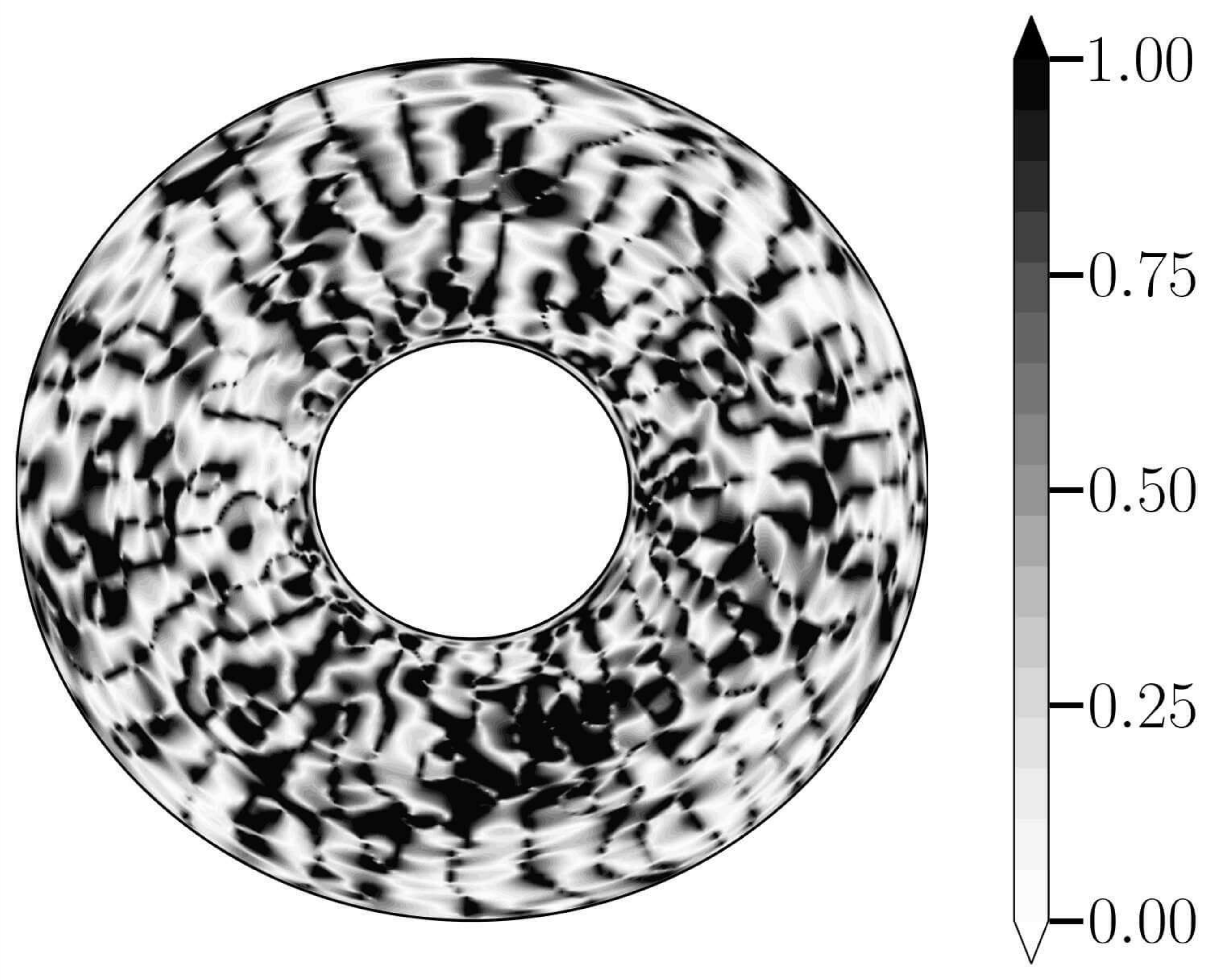}}\\ 
		\subfloat[\hspace{1.5cm} A \label{fig:MAC_b}]
		{\includegraphics[width=0.7\linewidth]{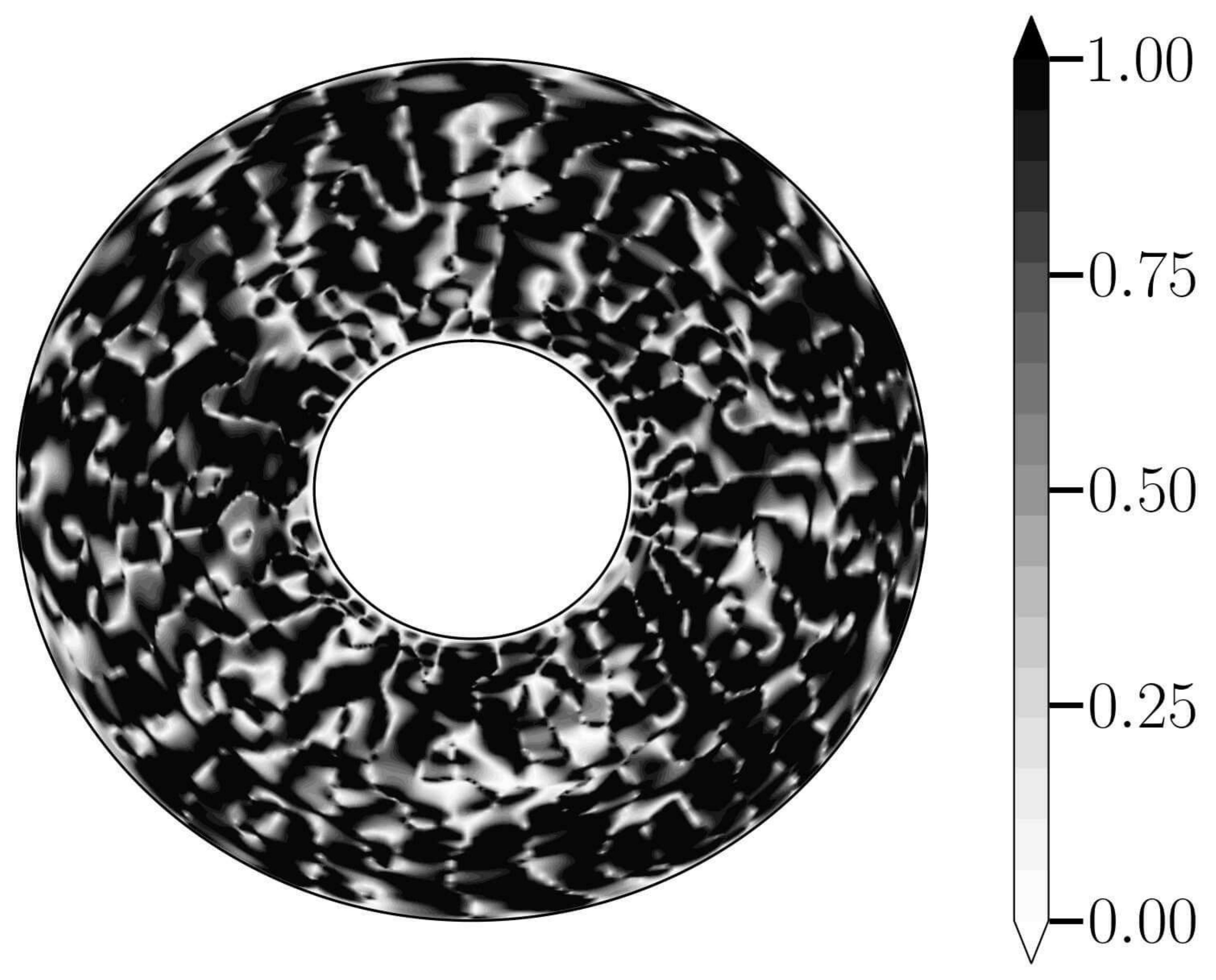}}\\
		\subfloat[\hspace{1.54cm} C  \label{fig:MAC_c}]
		{\includegraphics[width=0.7\linewidth]{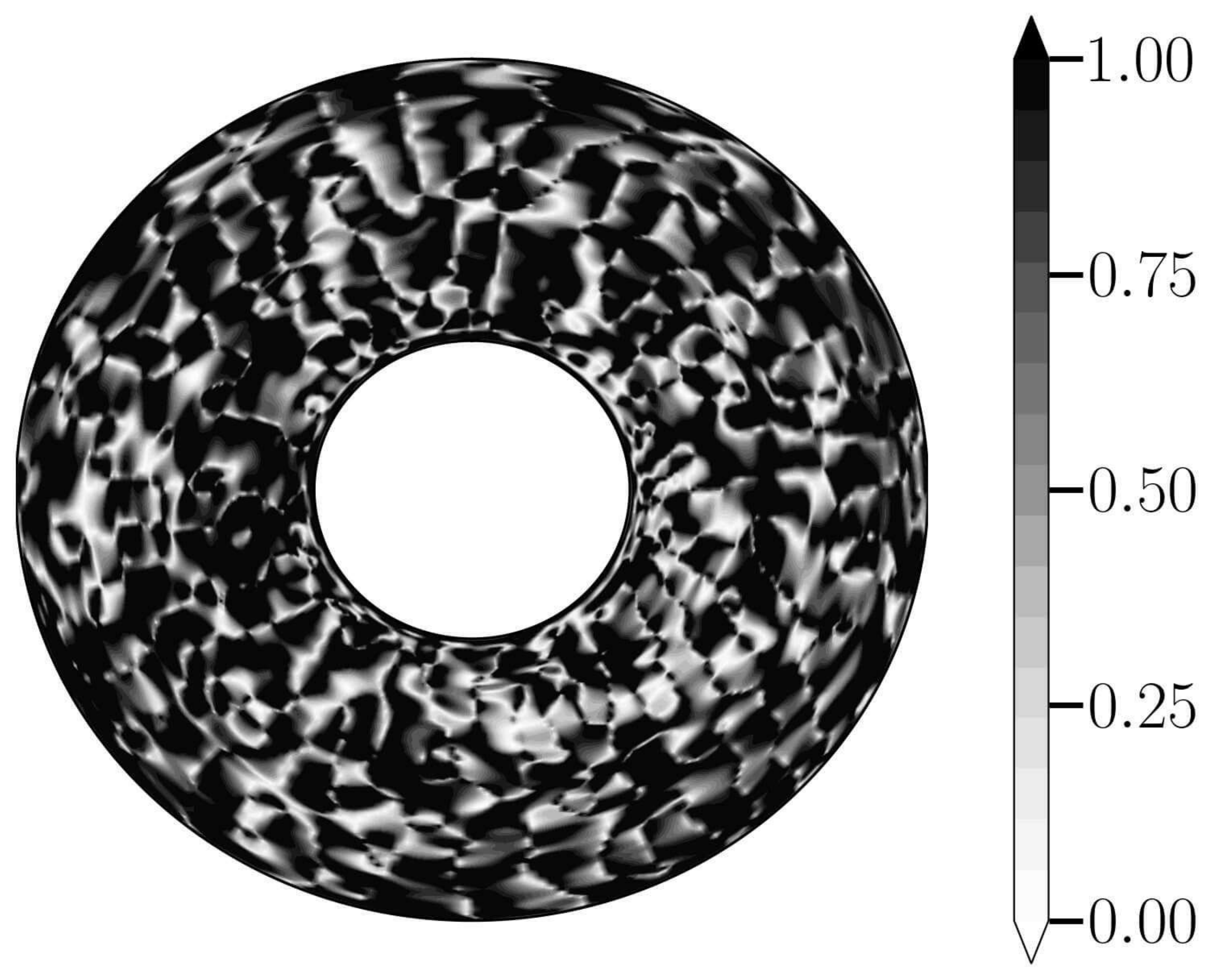}}  
	\end{minipage}
	\hspace*{-1cm}
	\begin{minipage}{0.45\linewidth}
		\centering
		\subfloat[\hspace{1.5cm} M 	\label{fig:MAC_d}]{\includegraphics[width=0.7\linewidth]{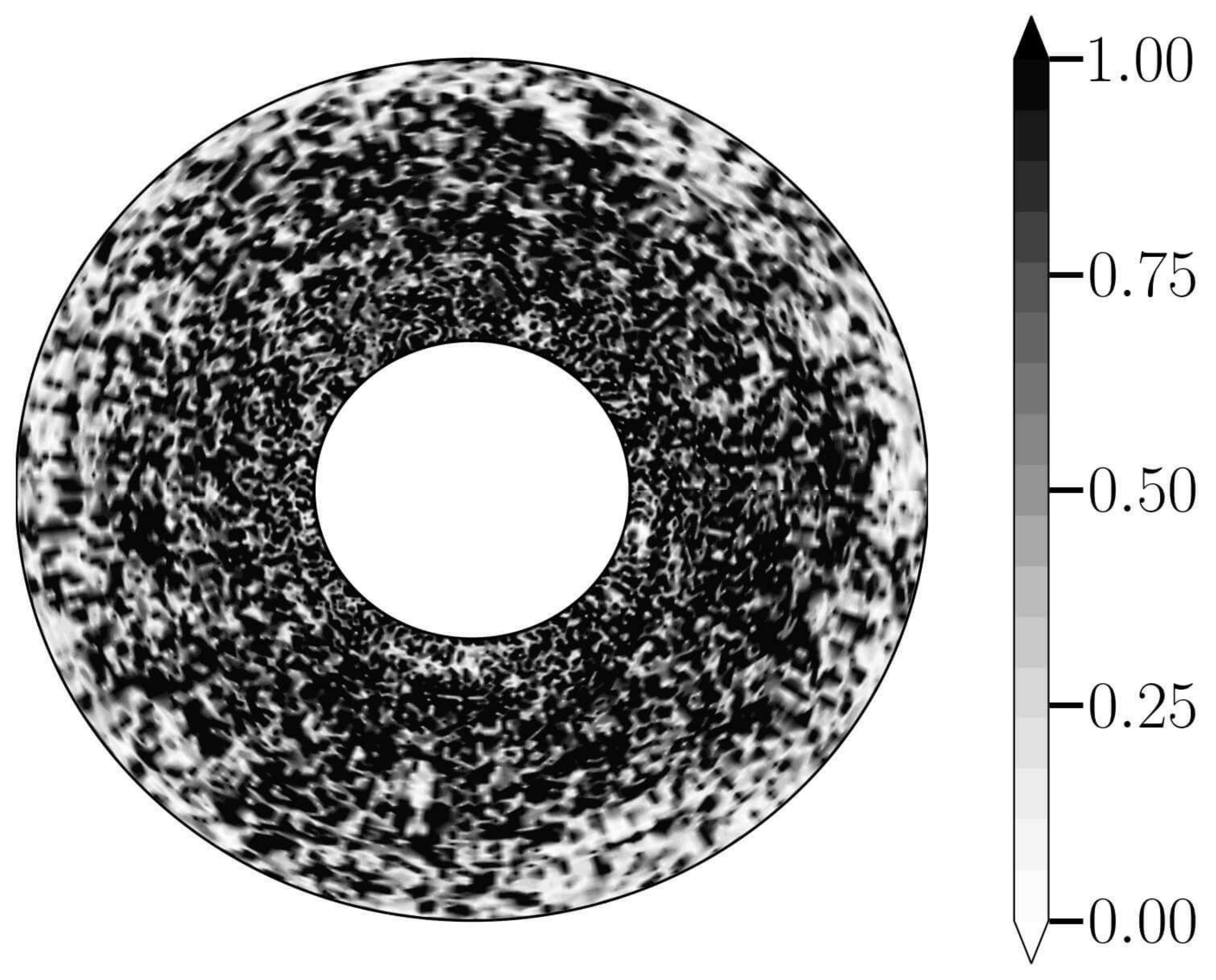}} \\
		\subfloat[\hspace{1.5cm} A 	\label{fig:MAC_e}]{\includegraphics[width=0.7\linewidth]{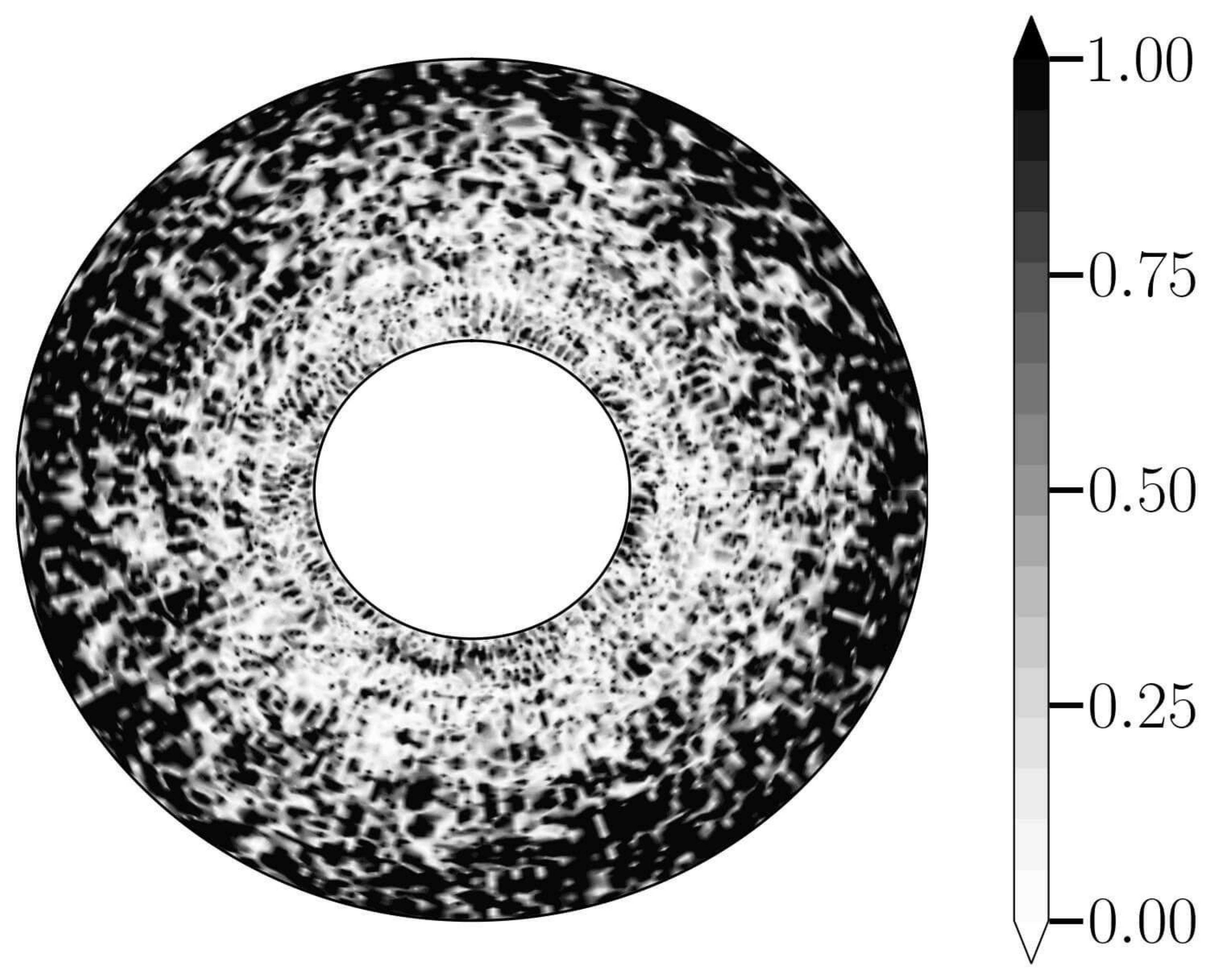}}\\
		\subfloat[\hspace{1.5cm} C 	\label{fig:MAC_f}]{\includegraphics[width=0.7\linewidth]{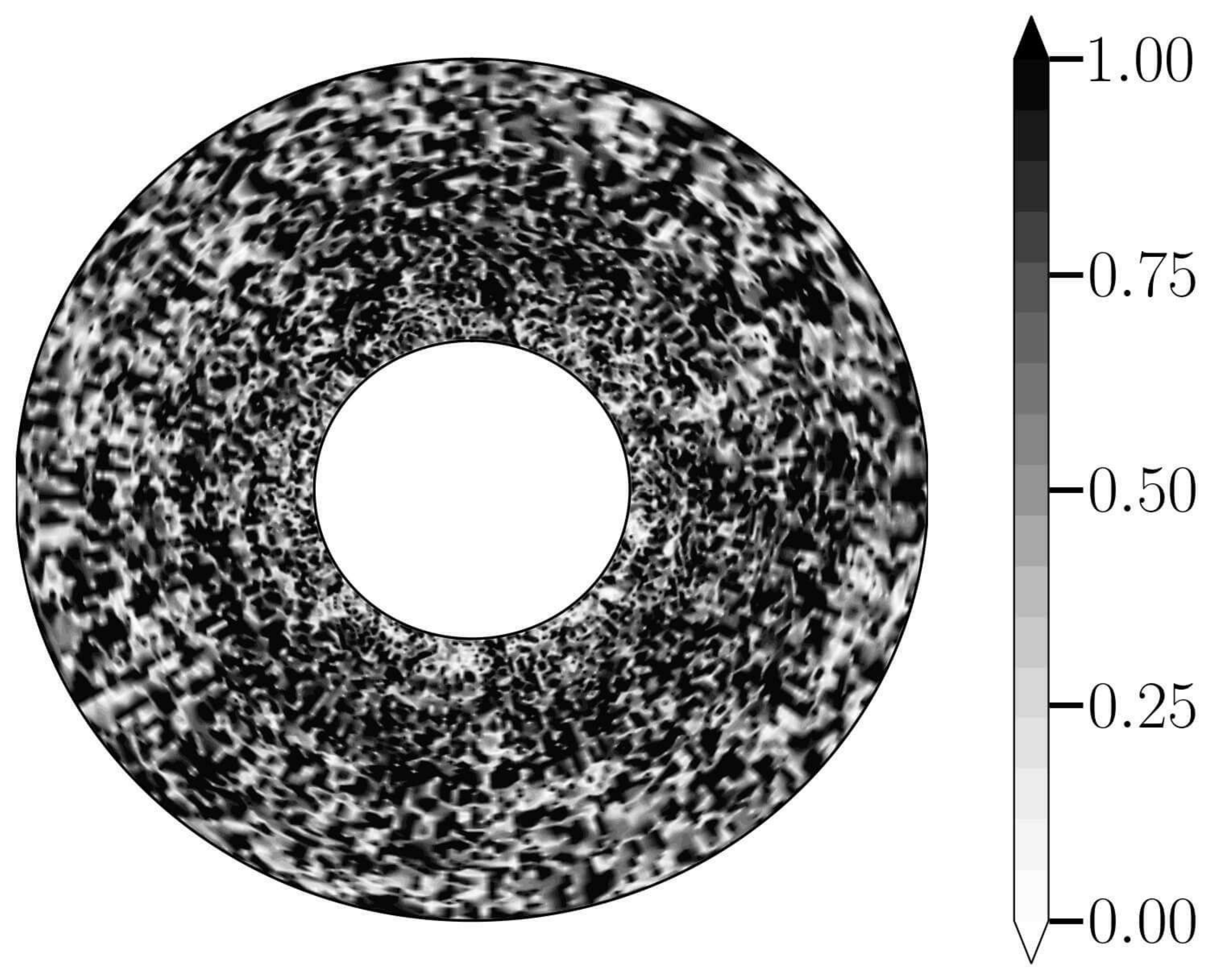}}				
	\end{minipage}
	\caption{ The ratio of magnitudes of the magnetic Lorentz
	(M), buoyancy (A) and Coriolis (C) force
	terms  in the $z$-vorticity 
		equation to the magnitude of the
		largest force among them, is 
		plotted on the horizontal section
		 $z=0.1$ for two ranges of scales 
		$l \leq 31$ in ((a)-(c)) and $l >100$ in ((d)-(f)).
		The model parameters are
		$Ra=400$, $Pm=Pr=1$, $E=1.2 \times 10^{-6}$.}
	\label{fig:MAC}
}
\end{figure}

	In the dynamo simulation at $E=1.2 \times 10^{-6}$
	and $Ra=400$, the 
	Coriolis and buoyancy forces are 
	in approximate balance for the relatively large 
	scales $l \leq 31$ (figure
	\ref{fig:MAC} (b) and (c)). However, as seen
	in figure \ref{fig:MAC}(a), 
	the Lorentz forces become significant
	in patches and  
	balance the Coriolis forces. Therefore, 
	local excitation of slow MAC
	waves in these scales is conceivable.
	 For the small scales of $l>100$,
	the buoyancy forces are restricted to the outer
	periphery of the shell (figure \ref{fig:MAC} (e)).
	 The dominant balance in these scales
	 is between the Lorentz and Coriolis forces 
	 (figure \ref{fig:MAC} (d) and \ref{fig:MAC} (f)). 
	In either range of harmonic degrees, 
	the nonlinear inertial and viscous forces
	are small compared with the other forces
	 in the bulk of the volume, and hence
	not shown.
	
	Before discussing the role of the slow MAC waves in
	the dynamo, we examine the relative magnitudes
	of the fundamental frequencies and show that wave motions
	correlate with helicity generation and dipole
	formation in the
	energy-containing scales.
	
\section{MAC waves, helicity and dipole formation}
		In the absence of magnetic diffusion,
the roots of the following 
characteristic equation \citep{07bussechapter, jfm21} give
the frequencies of waves produced in a rotating stratified
fluid layer subject to a magnetic field:	
	\begin{equation}
	(\omega^2 - \omega_{M}^2 -\omega_{A}^2)
	(\omega^2 -\omega_{M}^2)-\omega_{C}^2\omega^2 = 0,
	\label{chareqn}
	\end{equation}  
where the fundamental frequencies $\omega_M$, $\omega_A$ and $\omega_C$
represent Alfv\'en waves, internal gravity waves and linear inertial
waves respectively. In unstable stratification that drives planetary
core convection, $\omega_A^2 <0$, where $|\omega_A|$
 is simply a measure of the 
strength of 
buoyancy. The dimensionless 
values of $\omega_M^2$, $-\omega_A^2$ and
$\omega_C^2$ in the dynamo are given by 
\begin{equation}
\omega_M^2 = \frac{Pm}{E} (\bm{B} \cdot {\bm k})^2, \quad 
-\omega_A^2 = \frac{Pm^2 Ra}{Pr \,E} \, \left(\frac{{k_z}^2
+ m^2}{k^2}\right), \quad
\omega_C^2 = \frac{Pm^2}{E^2} \frac{k_z^2}{k^2},
\label{om2}
\end{equation}	
where $k_s$, $m$ and $k_z$ are the radial, azimuthal and
axial wavenumbers in cylindrical
coordinates $(s,\phi,z)$, and $k$ is the resultant
wavenumber. Here,
$\omega_A$ is evaluated on the
equatorial plane where the buoyancy force is maximum;
$\omega_M$ is based on the measured peak magnetic
field in the dynamo.
Using the scaling for time in the dynamo model, the
frequencies in \eqref{om2} are scaled
by $\eta/L^2$.

For the inequality $\omega_C > \omega_M > \omega_A$,
the roots of equation \eqref{chareqn},
\begin{eqnarray}
\omega_{f} 
= \pm\frac{1}{\sqrt{2}} \sqrt{\omega_A^2+\omega_C^2+2\omega_M^2+\sqrt{\omega_A^4+2\omega_A^2\omega_C^2+4\omega_M^2\omega_C^2+\omega_C^4}},\label{ndr1}\\
\omega_{s} = \pm\frac{1}{\sqrt{2}} \sqrt{\omega_A^2+\omega_C^2+2\omega_M^2-\sqrt{\omega_A^4+2\omega_A^2\omega_C^2+4\omega_M^2\omega_C^2+\omega_C^4}}.
\label{ndr2}
\end{eqnarray}
represent the fast ($f$) and slow ($s$) 
MAC waves. While the fast waves
are linear inertial waves weakly modified by the magnetic
field and buoyancy, the slow waves are magnetostrophic
\citep{67brag,73acheson,
07bussechapter,jfm21}.

\begin{figure}
	
	{
		\centering
		
		\subfloat[\label{fig:bp10_E6Ra400_a} ]{\includegraphics[width=0.45\linewidth]{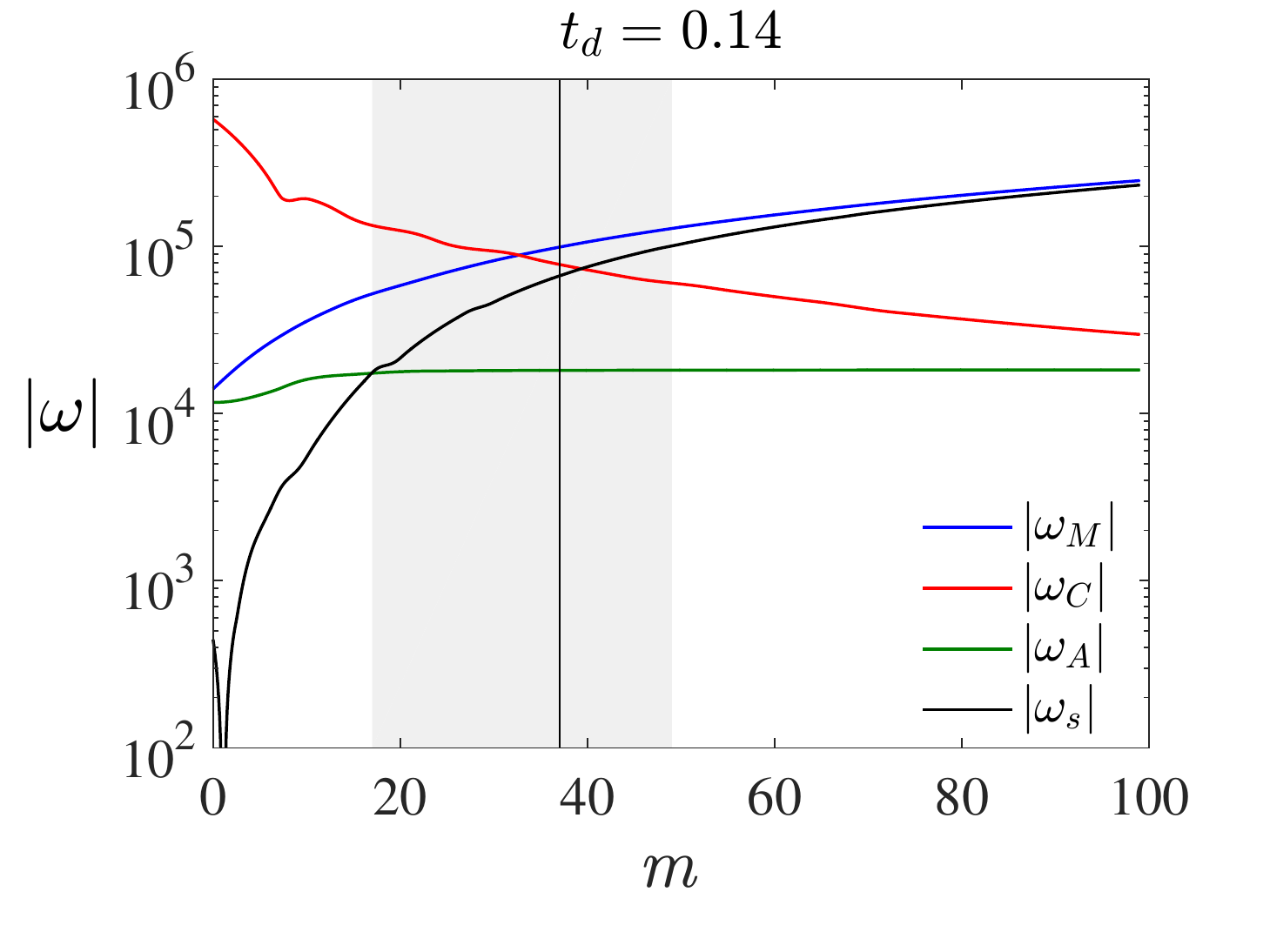}}		
		\subfloat[\label{fig:bp10_E6Ra400_b} ]{\includegraphics[width=0.45\linewidth]{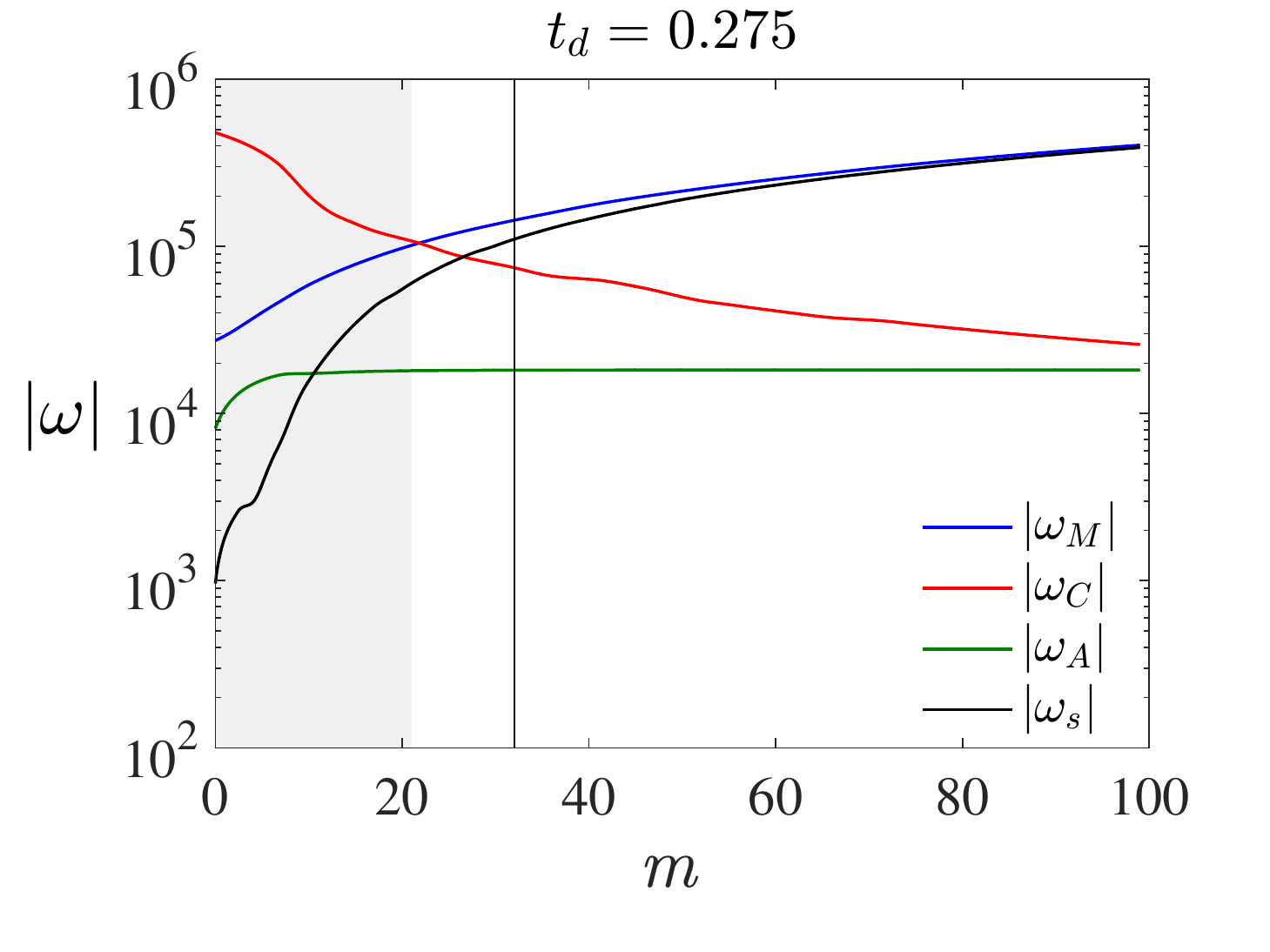}}	\\
		\vspace{-1cm}
		\subfloat[	\label{fig:bp10_E6Ra400_c} ]{\includegraphics[width=0.45\linewidth]{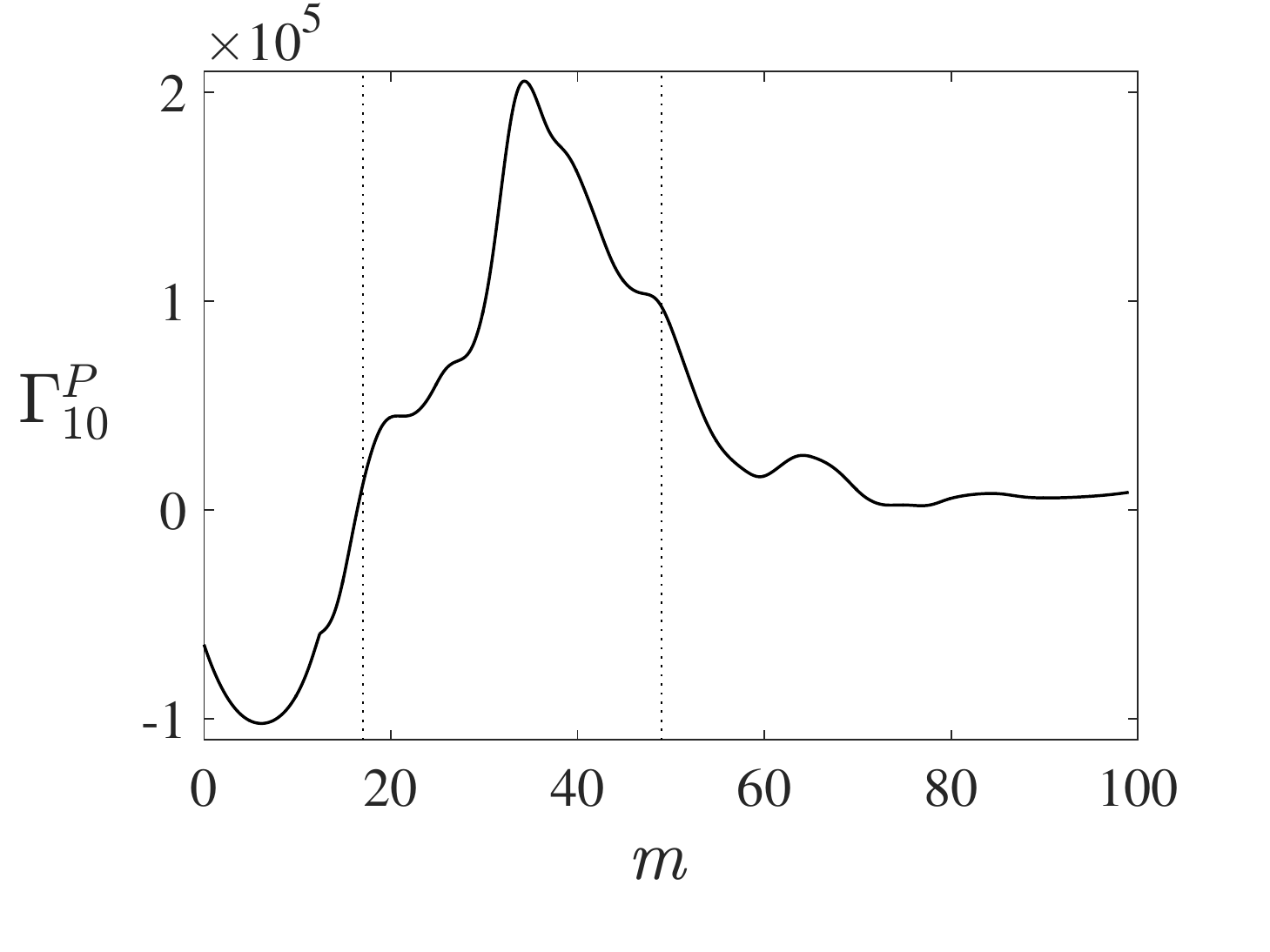}}		
		\subfloat[ 	\label{fig:bp10_E6Ra400_d}]{\includegraphics[width=0.45\linewidth]{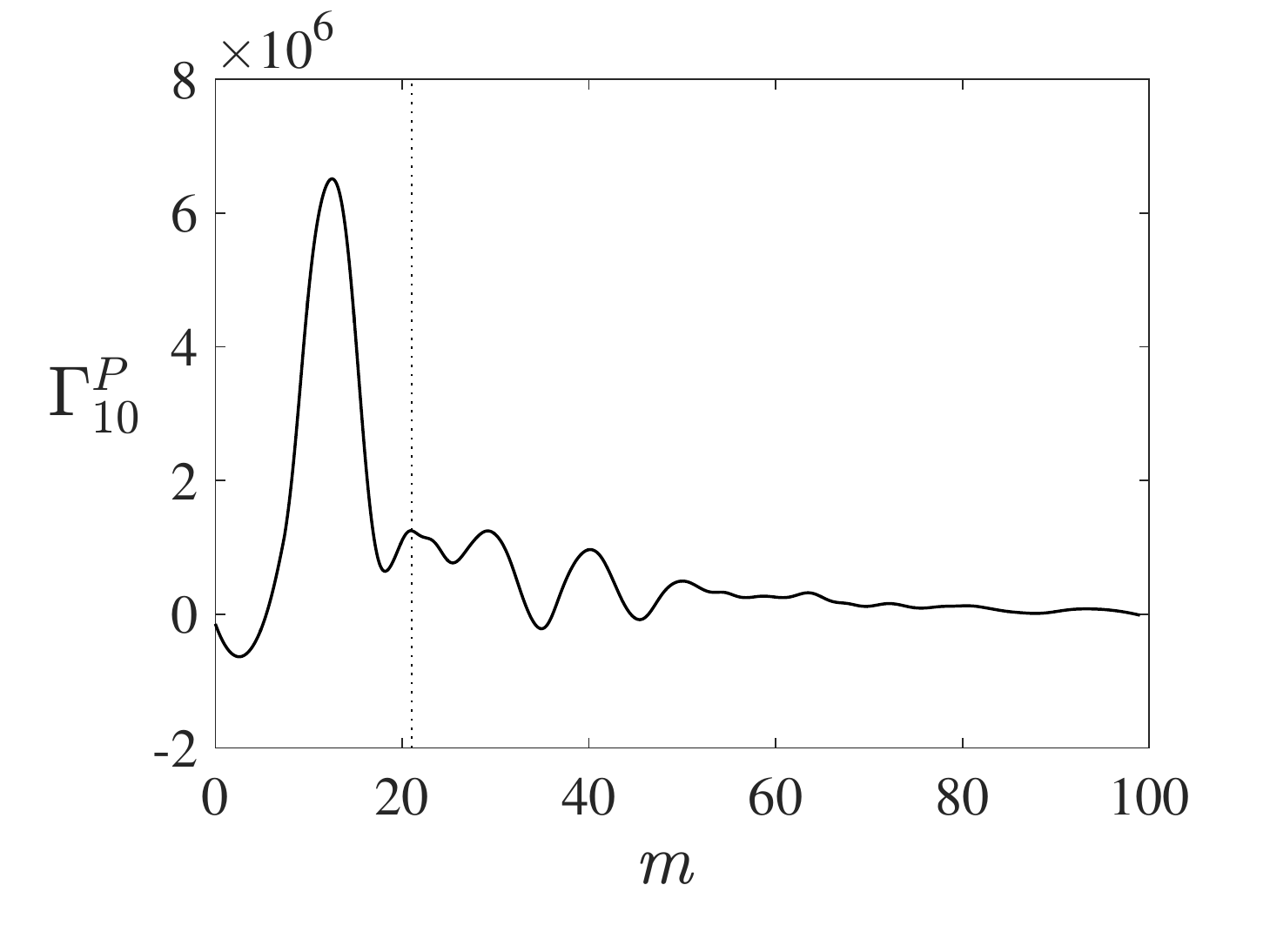}}\\	
		
		\caption{(a) \& (b): Absolute values of frequencies 
			 plotted for two snapshots of time during the evolution
			 of the dynamo from a small seed field magnetic field.
			The magnitudes of the following frequencies are shown:
			$\omega_C$ (linear inertial wave), 
			$\omega_M$ (Alfv\'en wave), 
			$\omega_A$ (internal gravity wave) and 
			$\omega_s$ (slow MAC wave). Since
			$\omega_A^2 <0$ in unstable stratification,
			$\omega_A$ is simply a measure of the strength
			of buoyancy in the dynamo.
		    The shaded grey area shows the scales where
			helicity is generated in the dynamo simulation 
			relative to the
			nonmagnetic simulation. The thin solid vertical 
			line shows
			the mean wave number of energy injection.
		    (c) \& (d): Contribution to the axial dipole energy
		    per unit time,
			$\int_{V} \bm{B}^{P}_{10}.[\bm{ 
			\nabla} \times(\bm{u} \times \bm{B})]dV$
			plotted as a function of the spherical harmonic
			order, $m$.
			The vertical dotted lines show the same range of scales 
			as in (a) \& (b), where helicity is generated.
			The dynamo parameters are 
			$Ra=400$, $Pm=Pr=1$, $E=1.2 \times 10^{-6}$.}
		\label{fig:bp10_E6Ra400}		
		
	}	
\end{figure}

\begin{figure}
		\centering	
		\includegraphics[width=0.5\linewidth]
		{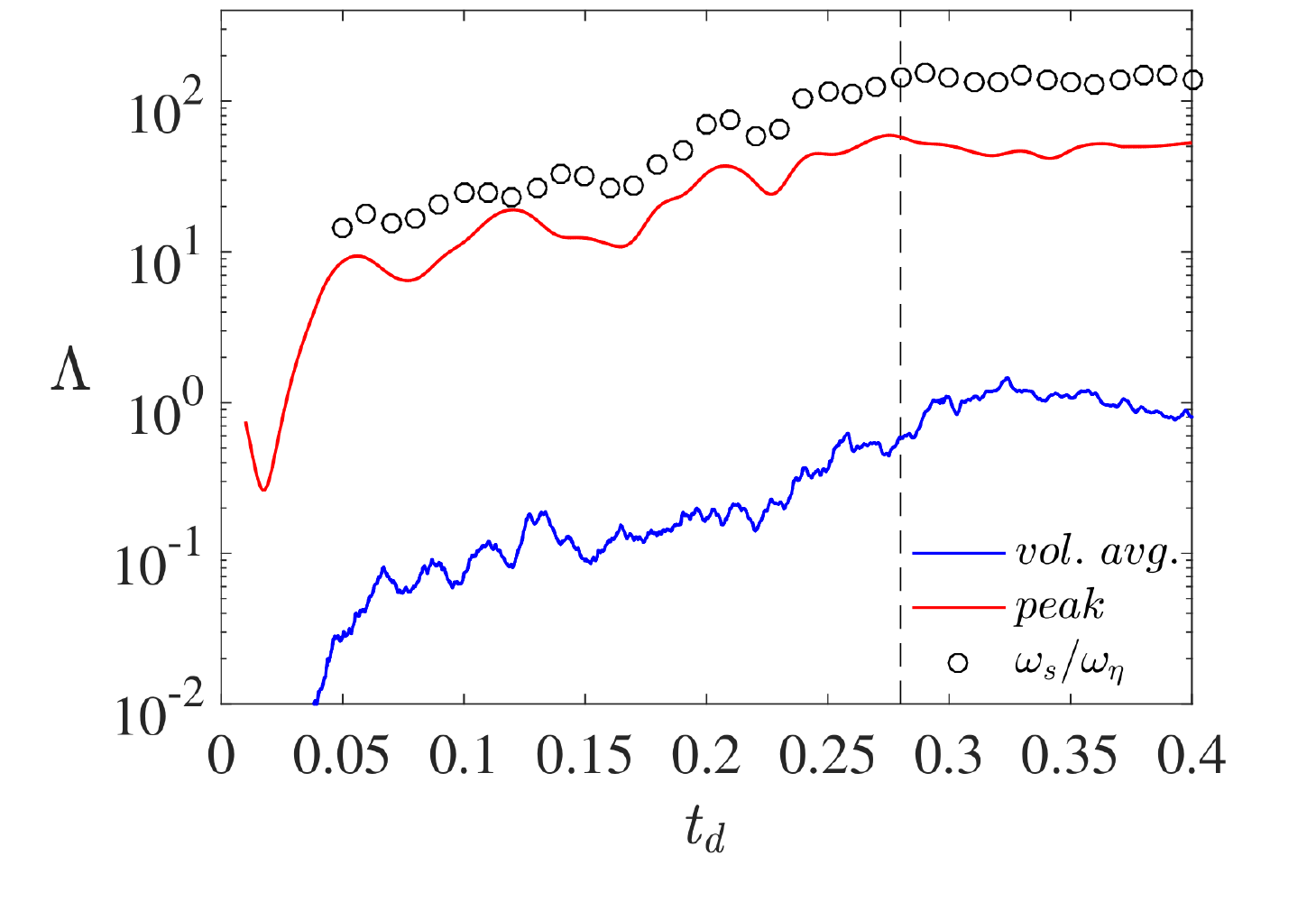}				
		\caption{Peak and
		volume-averaged values of the Elsasser
		number $\varLambda$ shown against magnetic
		diffusion time,
		starting from the initial seed field state
		to the saturated state of the dynamo.
		The symbols (circles) represent the instantaneous values
		of $\omega_s/\omega_\eta$, where $\omega_s$ is the 
		slow MAC wave frequency and $\omega_\eta$ is the
		magnetic diffusion frequency. The dashed vertical
		line marks the dipole formation time.
		The dynamo simulation has the parameters
		$Ra=400$, $Pm=Pr=1$, $E=1.2 \times 10^{-6}$.
		}
		\label{fig:els_peak}
	\end{figure}

In figure \ref{fig:bp10_E6Ra400}(a) and (b), the
magnitudes of the fundamental frequencies are shown
as a function of the azimuthal wavenumber $m$. Two
times are analysed 
in the growth phase of the dynamo run at
$E=1.2 \times 10^{-6}$, $Ra=400$ and $Pr=Pm=1$.
The frequencies are computed from \eqref{om2} using
the mean values of the $s$ and $z$ wavenumbers. For
example, real space integration over $(s,\phi)$
gives the kinetic energy as a function of $z$, the Fourier
transform of which gives the one-dimensional spectrum
$\hat{u}^2 (k_z)$. Subsequently, we obtain
\begin{equation}
\bar{k}_z = \frac{\Sigma k_z \, \hat{u}^2 (k_z)}
{\Sigma \hat{u}^2 (k_z)}.
\label{kzmean}
\end{equation}
A similar approach gives $\bar{k}_s$. The computed
frequencies in figure \ref{fig:bp10_E6Ra400}(a) and (b)
satisfy the inequality $|\omega_C| > |\omega_M| >
|\omega_A|$ in the energy-containing scales of the
dynamo spectrum, indicating that the
MAC waves
would be generated in these scales. We emphasize
that this inequality would be obtained only if the
measured peak magnetic field is used in the evaluation of
$\omega_M$, corresponding to local Elsasser
numbers $\varLambda \gg 1$
(see \cite{jfm21} and figure \ref{fig:els_peak} in
Section \ref{concl}).
The range of
scales with the above frequency 
inequality narrows as the field intensity increases
in time, and close to dipole formation
($t_d=0.275$), this inequality is confined
to wavenumbers $m<19$. The scales of helicity
generation, shown in shaded grey bands, are
obtained from the differences between the helicity
spectra of the magnetic and 
equivalent non-magnetic runs. Notably, the region of
helicity generation overlaps with the scales where the
MAC waves are generated. The slow MAC wave frequency 
merges with the Alfv\'en wave frequency at large
$m$, where $\omega_M$ is the dominant frequency.

The contribution of convection to the axial
dipole field energy per unit time is given
by \citep[e.g.][]{buffett2002}
\begin{equation}
\Gamma^{P}_{10}= \int_{V} \bm {B}^{P}_{10} \cdot
[\mathbf \nabla \times(\bm {u} \times \bm{ B})]dV.
\label{gamma10}
\end{equation}
The spectral distribution of $\Gamma^{P}_{10}$ is shown in figure
\ref{fig:bp10_E6Ra400} (c) and (d). 
Evidently, the maximum contribution to the dipole energy 
	occurs in the scales where helicity is generated by the
	magnetic field. The strong correlation between MAC wave
	formation, helicity generation, and in turn, the axial dipole
	field energy, is also noted in the dynamo
	simulation at $E=10^{-5}$.
 
\begin{figure}
		\centering
		
		\subfloat[	\label{fig:freq_hel_a}]{\includegraphics[width=0.5\linewidth]{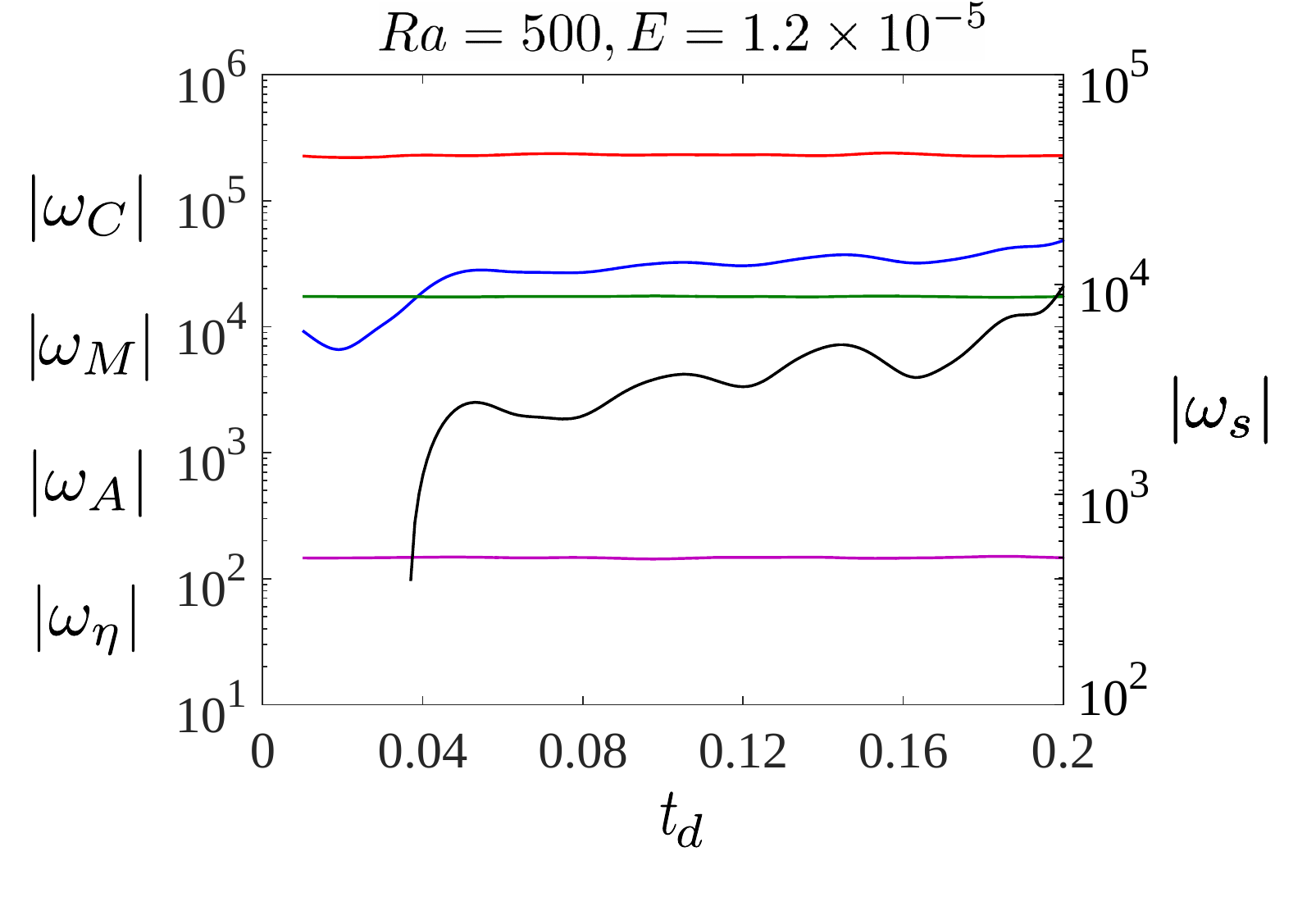}}	
		\subfloat[	\label{fig:freq_hel_b}]{\includegraphics[width=0.5\linewidth]{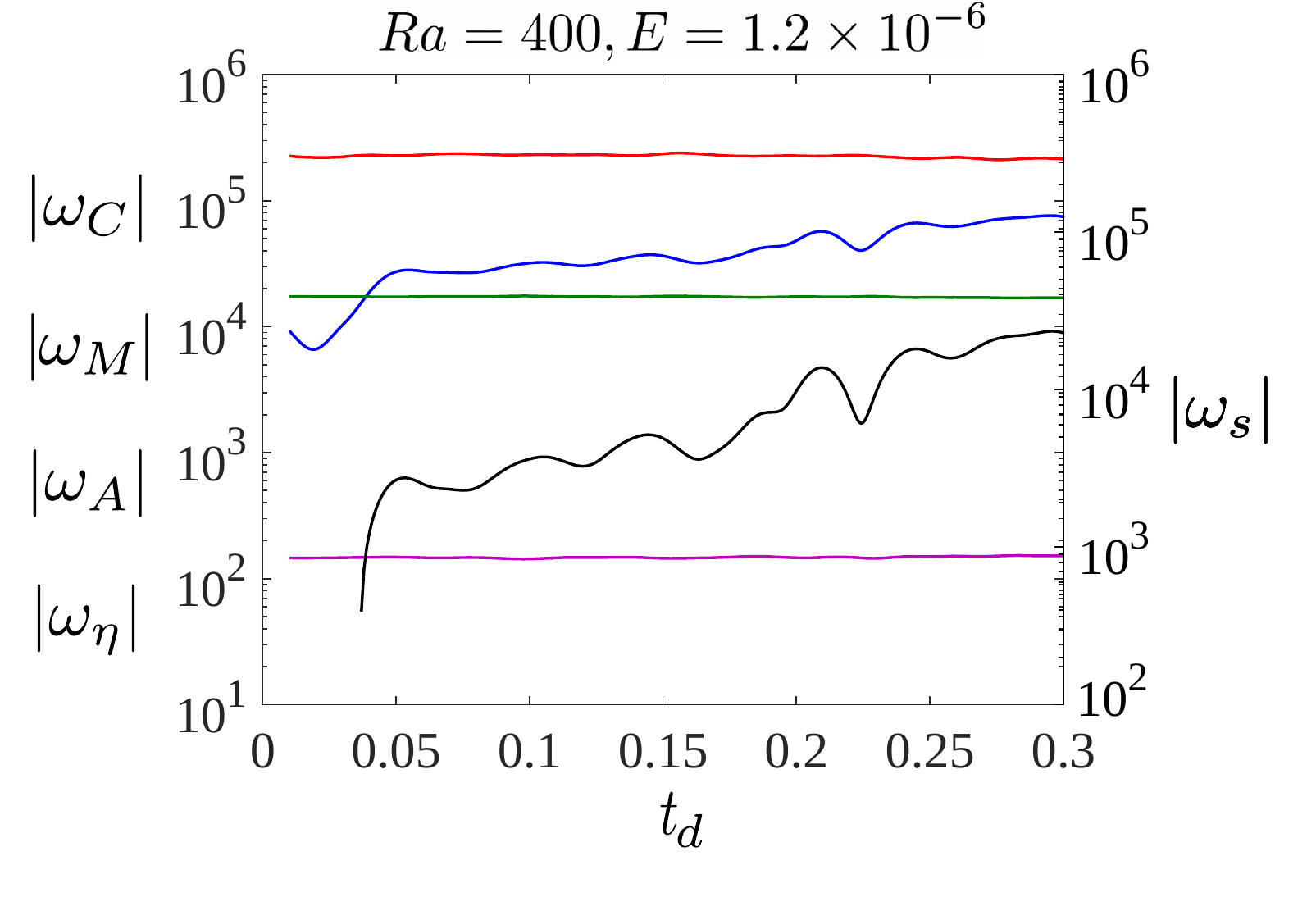}}\\
			
		\caption{Absolute values of the dynamo frequencies
			plotted against time (in units
			of the magnetic diffusion time $t_d$).
			Both the 
			simulations study the evolution of the dynamo
			starting from a small seed magnetic field.
			The frequencies are calculated at the 
			mean azimuthal wavenumber ($\bar{m}=$10 for (a) and
			11 for (b)) of the range of
			scales where MAC waves are active at dipole
			formation time. The
		     axial dipole forms from a multipolar state at
		     $t_d \approx 0.125$ in (a) and $t_d \approx 0.28$
		     in (b).
			The frequencies shown (with line colours in 
			brackets) are as follows:
		     $\omega_M$ (blue), $\omega_C$ (red), $\omega_A$ (green),
		     $\omega_{\eta}$ (magenta), $\omega_s$ (black). }		     
		\label{fig:freq_hel}
\end{figure}

	Figure \ref{fig:freq_hel}(a) and (b) shows the 
	fundamental frequencies and the slow MAC wave
	frequency plotted against time for two dynamo
	simulations that begin from a small seed magnetic field.
	The frequencies are calculated at
	mean azimuthal wavenumber of the range of scales where
	the slow waves are thought to be active
	at dipole formation time. The dimensionless 
	magnetic diffusion frequency $\omega_\eta$ is
	given by $(\bar{k} L)^2$. It is evident that
	the frequency inequality $|\omega_C| > |\omega_M|
	> |\omega_A| > |\omega_\eta|$ is satisfied at times
	approaching axial dipole formation and beyond.
	The slow MAC waves, whose
	frequency $\omega_s$ is given by the black solid line, are
	generated for $|\omega_M| > |\omega_A|$, since
	\citep{67brag,07bussechapter}
	\begin{equation}
	\omega_s \approx \frac{\omega_M^2}{\omega_C} \left(1+\frac{\omega_A^2}
	{\omega_M^2} \right)^{1/2},
	\label{wmacapprox}
	\end{equation}
	where $\omega_M^2/\omega_C$ is the Magneto-Coriolis (MC)
	wave frequency.

	\begin{table}
	\centering
	\begin{tabular} {l l l | l l l}
	\hline
	\multicolumn{3}{c|}{$E= 1.2 \times 10^{-5}$, $Ra=500$} &
        \multicolumn{3}{c}{$E= 1.2 \times 10^{-6}$, $Ra=400$}\\
        $t_d$ & $\omega_M/\omega_C$ & $Le$  & $t_d$ & $\omega_M/\omega_C$ & $Le$ \\
        \hline
        0.033 & 0.108 & 0.005 & 0.041 & 0.076 & 0.003 \\
        0.04 & 0.147 & 0.006 & 0.14 & 0.162 & 0.0065 \\
        0.125 & 0.310 & 0.014 & 0.28 & 0.345 & 0.014 \\
        \hline
        \end{tabular}
        \caption{Comparison of the values of the Lehnert number
       $Le$ and the frequency ratio $\omega_M/\omega_C$ at three
       points in time (in units of the magnetic diffusion time $t_d$)
       during the growth phase of two dynamo models.
       The times considered are those at incipient slow MAC
       wave generation, onset of helicity generation, and
       axial dipole formation.
       The evolution in time of the measured frequencies 
       in these models is shown in figure \ref{fig:freq_hel}(a)
       and (b).}
\label{lecomp}
        \end{table}

   The Lehnert number in the dynamo simulations, evaluated by
\begin{equation*}
	Le= |\bm{B}| \left(\frac{E}{Pm} \right)^{1/2} \, \frac{m}{2 \pi},   
	\end{equation*}	
	has its origin in $(\omega_M/\omega_C)_0$, the frequency ratio
at the initial state of a buoyant blob released into the flow \citep{jfm21}. As blobs
evolve in time into columns, the wavenumbers $k_s$ and $k_z$ decrease relative
to $m$, so the instantaneous value of $(\omega_M/\omega_C)$ is at least one
order of magnitude higher than $Le$ (table \ref{lecomp}). For values of $(\omega_M/\omega_C) \sim 0.1$,
the intensity of slow MAC wave motions would be comparable to that
of the fast waves \citep{jfm21}. Consequently, one would expect the helicity generated by the slow wave
motions to be of the same order of magnitude as that of the fast waves.
The approximately two-fold increase in the helicity as the dynamo evolves 
from a seed magnetic field (figure \ref{fig:hel_curvefit} (a) \&
(b)) suggests that the helicity generated by slow wave motions
in the dynamo may be comparable to that produced by the
fast inertial waves in nonmagnetic convection.

While the frequency diagrams in figure \ref{fig:bp10_E6Ra400}
and figure \ref{fig:freq_hel} suggest the active role of
slow MAC waves in dipole formation, conclusive evidence for
the existence of these waves necessitates
visualization of their propagation, 
which is presented in the following section.

\begin{figure}

{
	\centering
	
	\subfloat[\label{fig:GV_t0_a}] {\includegraphics[width=0.5\linewidth]
	{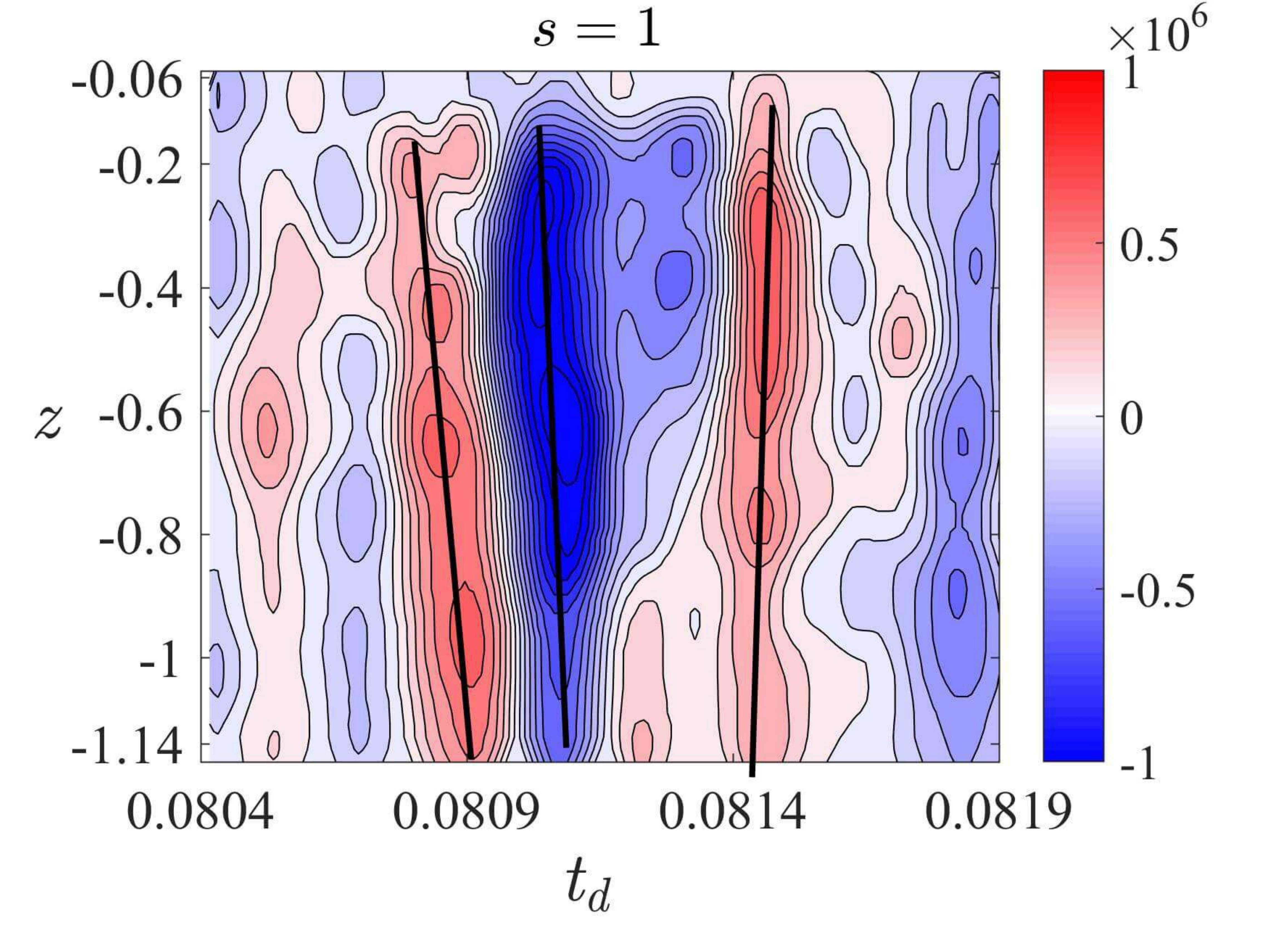}}
	\subfloat[\label{fig:GV_t0_b}] {\includegraphics[width=0.5\linewidth ]
	{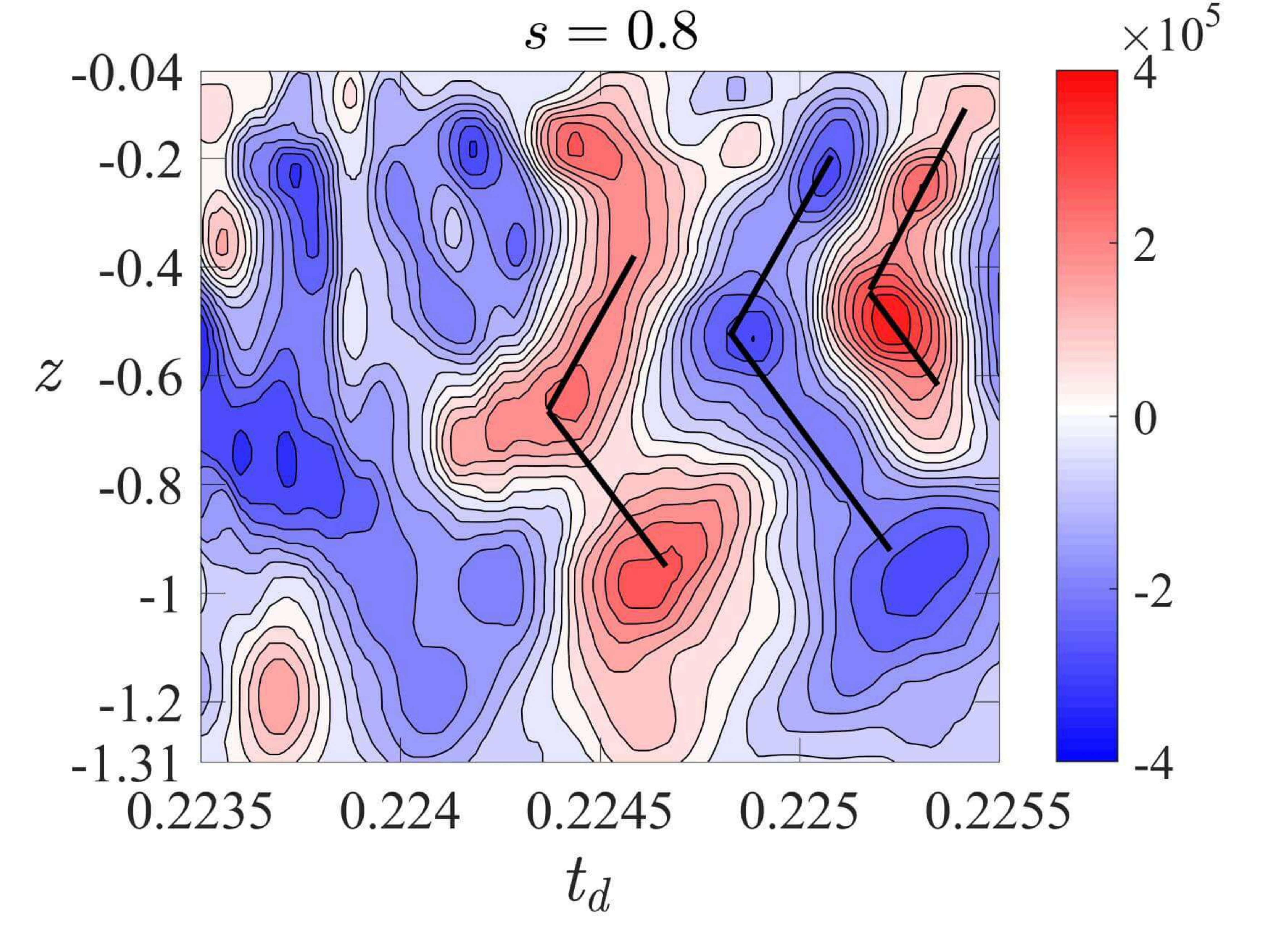}}\\
	\subfloat[\label{fig:GV_t0_c}] {\includegraphics[width=0.5\linewidth]
	{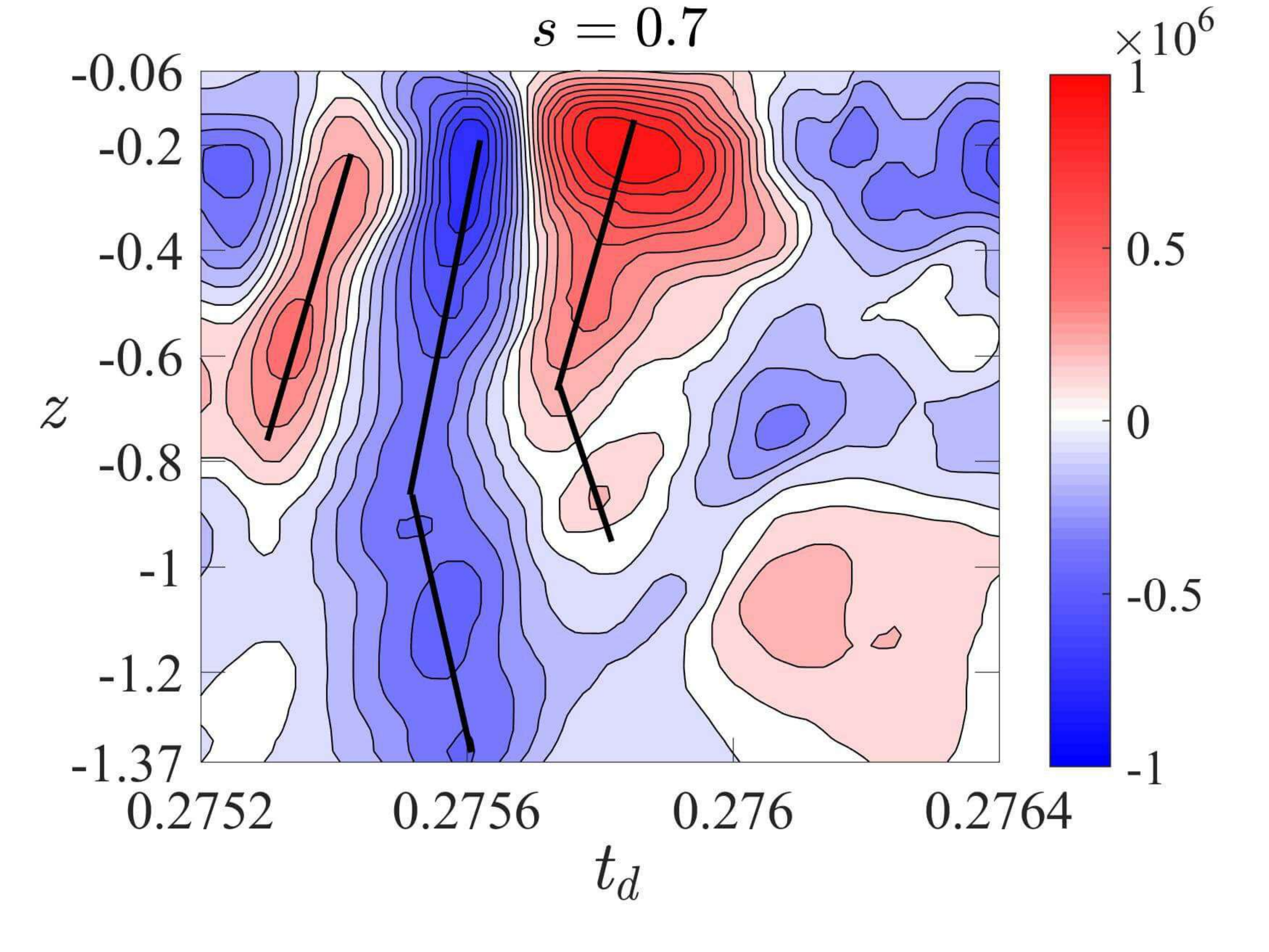}}
	\subfloat[\label{fig:GV_t0_d}] {\includegraphics[width=0.5\linewidth ]
	{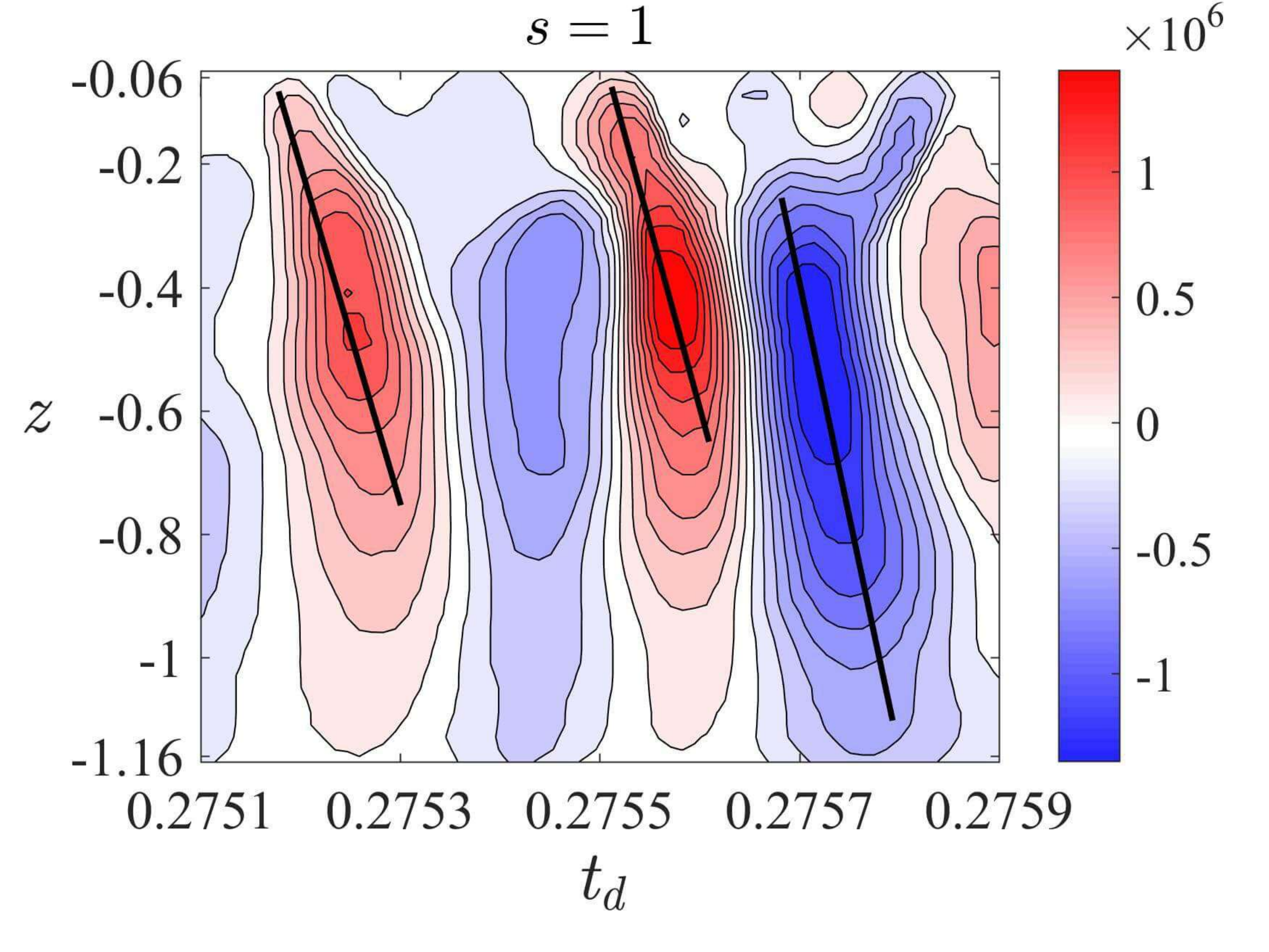}}
	
	\caption{(a) Contour plot of $\dot{u_{z}}$ for the time
		interval $t_{d}=$0.08--0.083 and 
		$l \leq 42$. (b) $\dot{u_{z}}$ for the time
		interval $t_{d}=$0.224--0.225 and 
		$l \leq 36$. (c) \& (d) $\dot{u_{z}}$ for the time
		interval $t_{d}=$0.274--0.278 and 
		$l \leq 31$. The cylindrical radius $s$ at which each
		plot is produced is given above the respective
		panel.
		The parameters of the dynamo simulation
		are $E=1.2 \times 10^{-6}$, $Ra=400$, $Pm=Pr=1$.
		The black lines indicate 
		the direction of travel of the waves and their 
		slope gives the group velocity. The estimated
		group velocity of the fast and slow MAC waves ($U_f$ and
		$U_s$ respectively)  and the measured group velocity
		$U_{g,z}$ are given in table
		\ref{tab:gv}.}
	\label{fig:GV_t0}

}

\end{figure}

\begin{figure}

{
	\centering
	\subfloat[	\label{fig:freq_butterfly_a}]{\includegraphics[width=0.45\linewidth]{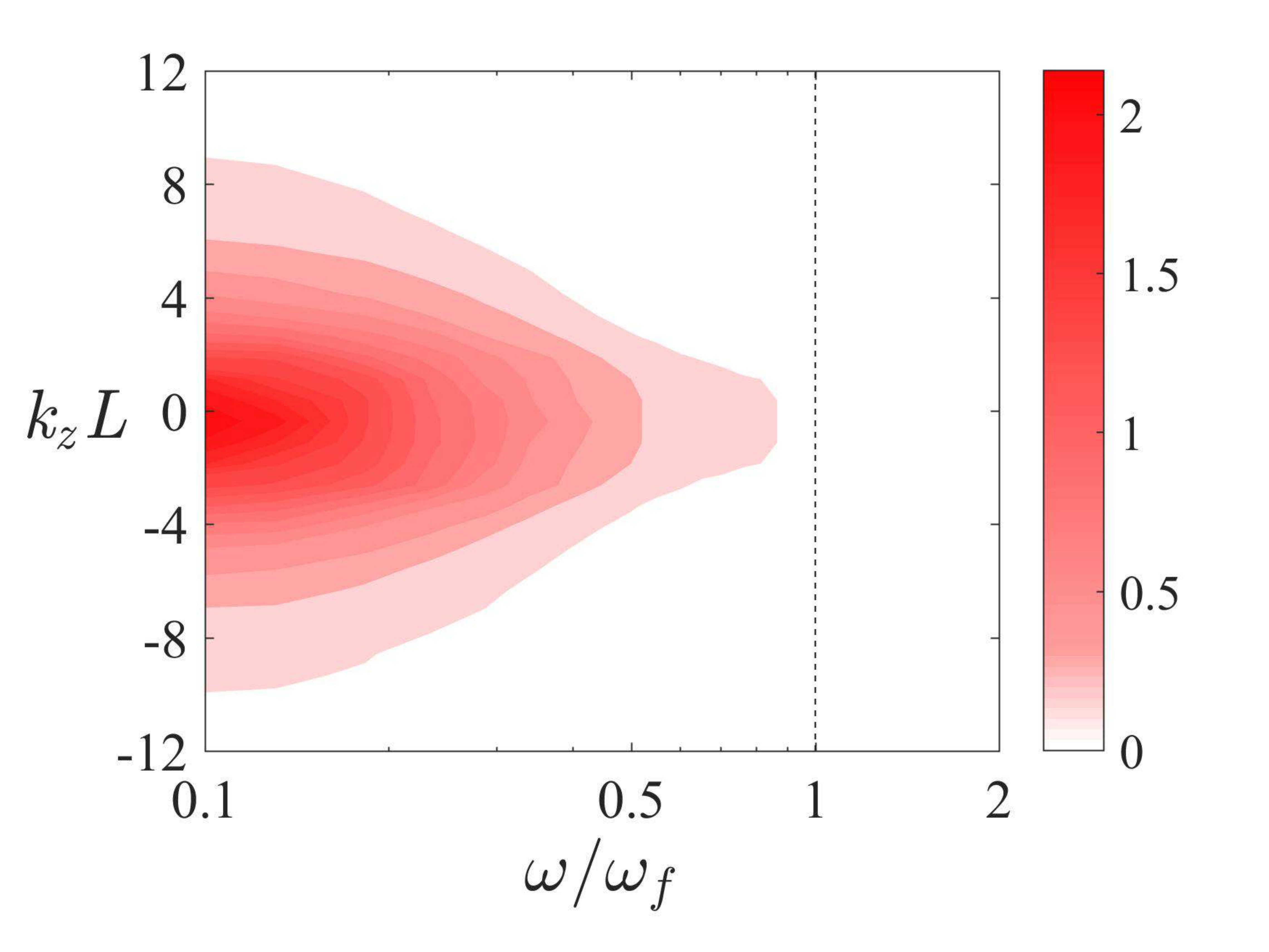}}
	\subfloat[\label{fig:freq_butterfly_b}]{\includegraphics[width=0.45\linewidth]{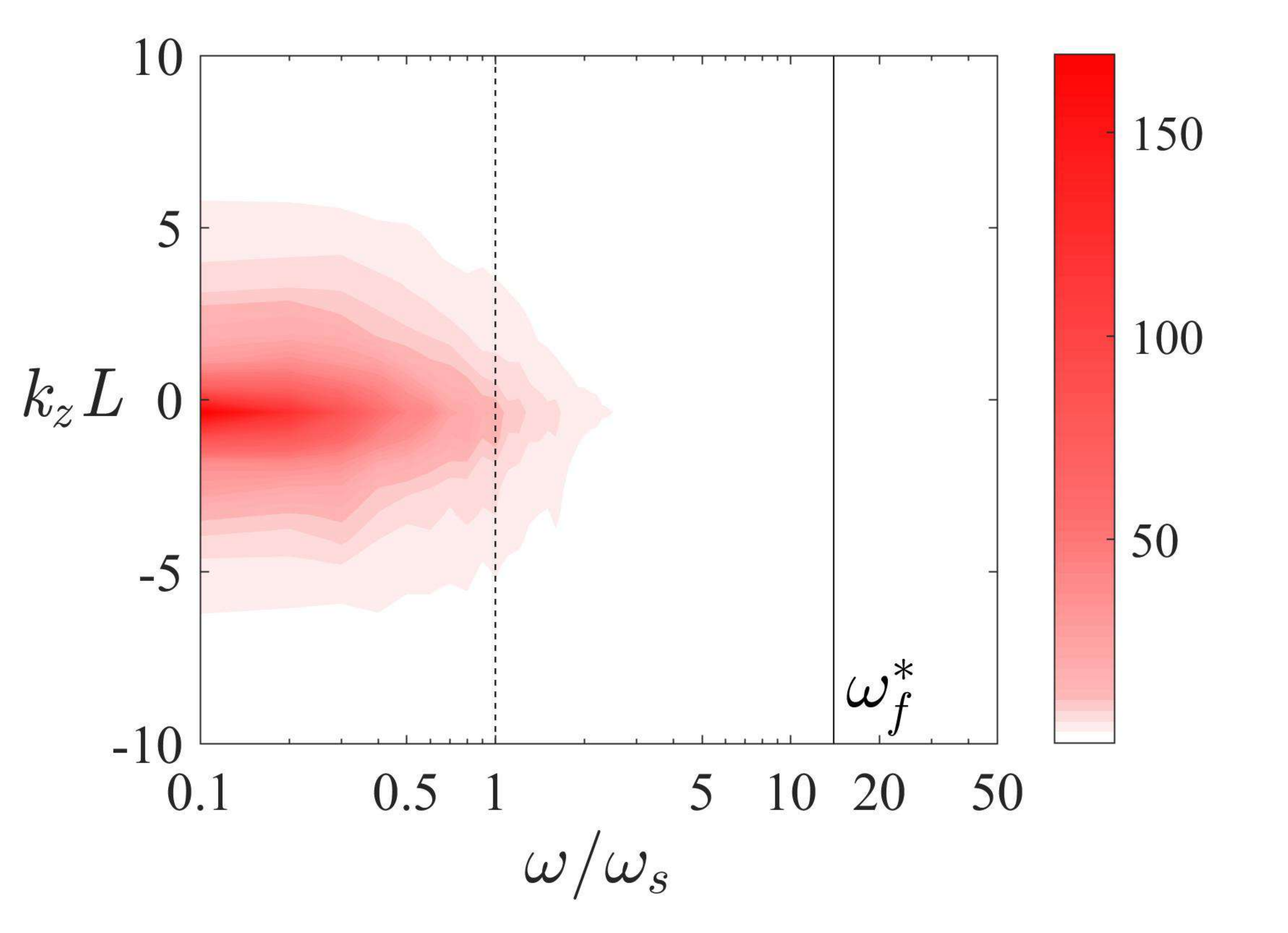}}\\
	
	\caption{(a) $\phi$-averaged FFT of $\dot{u_{z}}$ 
		at cylindrical radius 
		$s=0.7$ for the scales $l \leq 42$ in the time interval
		$t_{d}=0.08-0.082$. (b) $\phi$-averaged FFT of $\dot{u_{z}}$ 
		at $s=0.7$ for $l \leq 31$ in the time interval
		$t_{d}=0.274-0.278$. The range $l \leq l_E$
		narrows down as the field intensity increases
		with time. The dynamo parameters
		are $Ra=400$, $Pm=Pr=1$, $E=1.2 \times 10^{-6}$.
		The dotted vertical lines correspond to
		$\omega/\omega_{f}=1$ in (a) and
		$\omega/\omega_{s}=1$ in (b), where 
	    $\omega_{f}$ and $\omega_{s}$ are the estimated fast and
	    slow MAC wave frequencies. In (b), $\omega^{*}_{f}$ =
		$\omega_{f}/\omega_{s}$.}   
	\label{fig:freq_butterfly}
}                                                                                                                                                                              	
\end{figure}

\begin{figure}

{
	\centering
	
	\subfloat[	\label{fig:GV_mag_a}]{\includegraphics[width=0.5\linewidth]{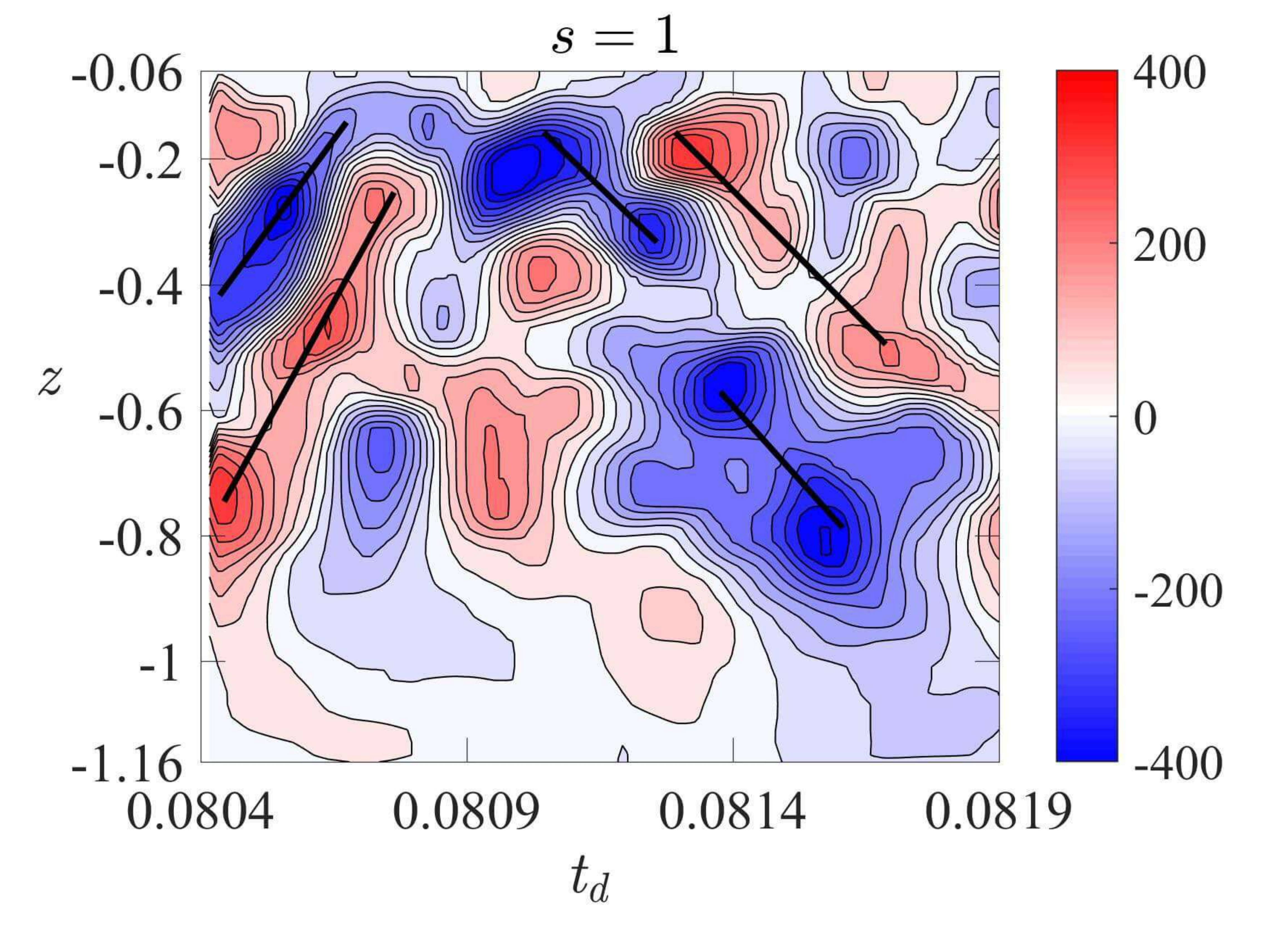}}		
	\subfloat[	\label{fig:GV_mag_b}]{\includegraphics[width=0.5\linewidth]{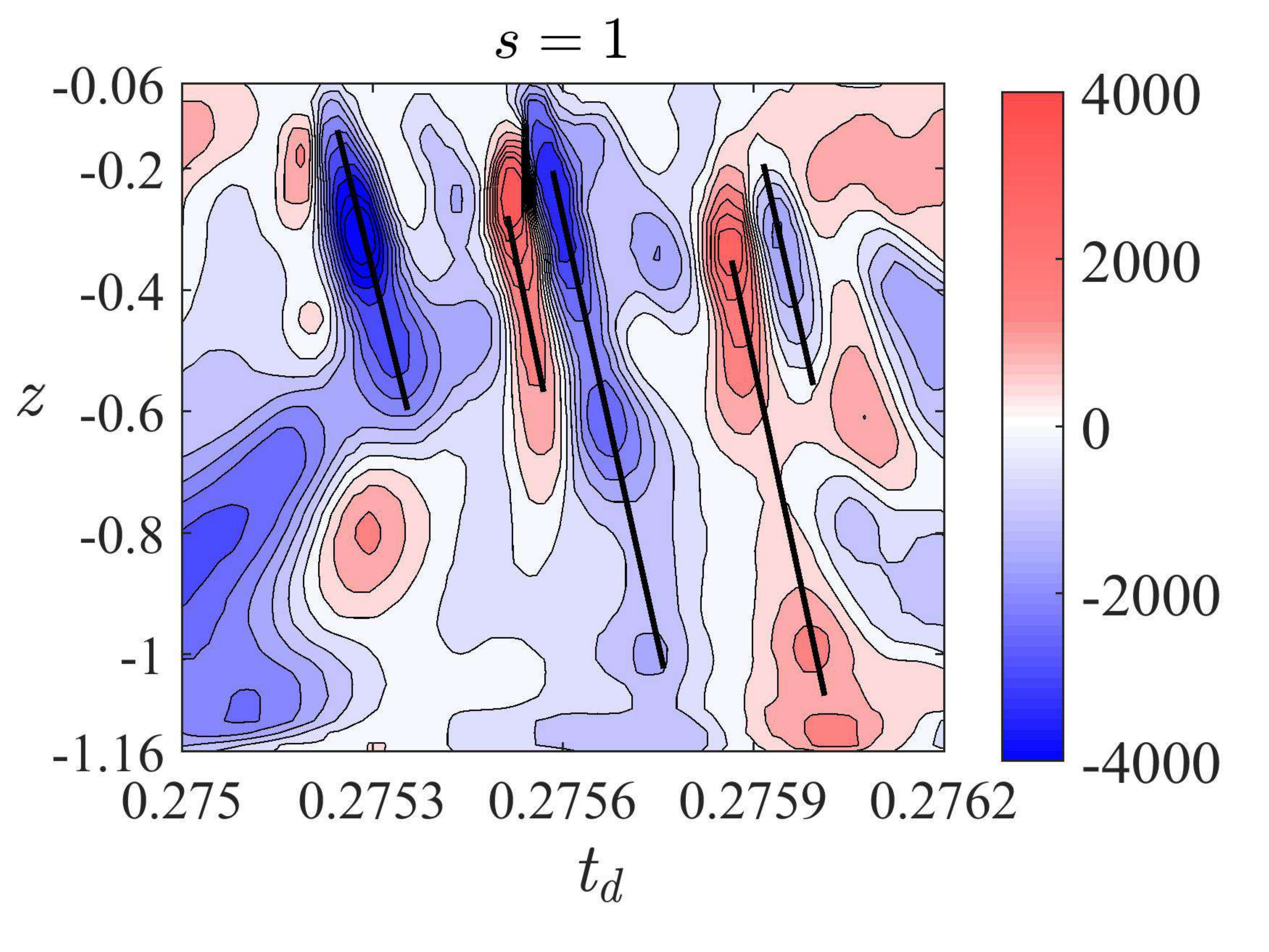}}

	\caption{Contour plots of $\dot{B_{z}}$ shown 
		for two time intervals.  (a) 
		$t_{d}=0.063-0.068$, $l \leq 42$. (b)
		$t_{d}=0.274-0.278$, $l \leq 31$.
		The cylindrical radius $s$ at which each
		plot is produced is given above the respective
		panel.
		The black lines indicate the direction of travel of 
		the wave and their slope gives the measured
		group velocity. The dynamo parameters
		are $Ra=400$, $Pm=Pr=1$, $E=1.2 \times 10^{-6}$.
		The estimated
		group velocity of the fast and slow MAC waves ($U_f$ and
		$U_s$ respectively)  and the measured group velocity
		$U_{g,z}$ are given in table
		\ref{tab:gv}.
		}
	
		\label{fig:GV_mag}
}

\end{figure}

\begin{figure}
{\centering
	
	\begin{minipage}[t]{0.5\linewidth}
		\centering
		\subfloat[	\label{fig:GV_Ra2000_a}]{\includegraphics[width=1\linewidth]
			{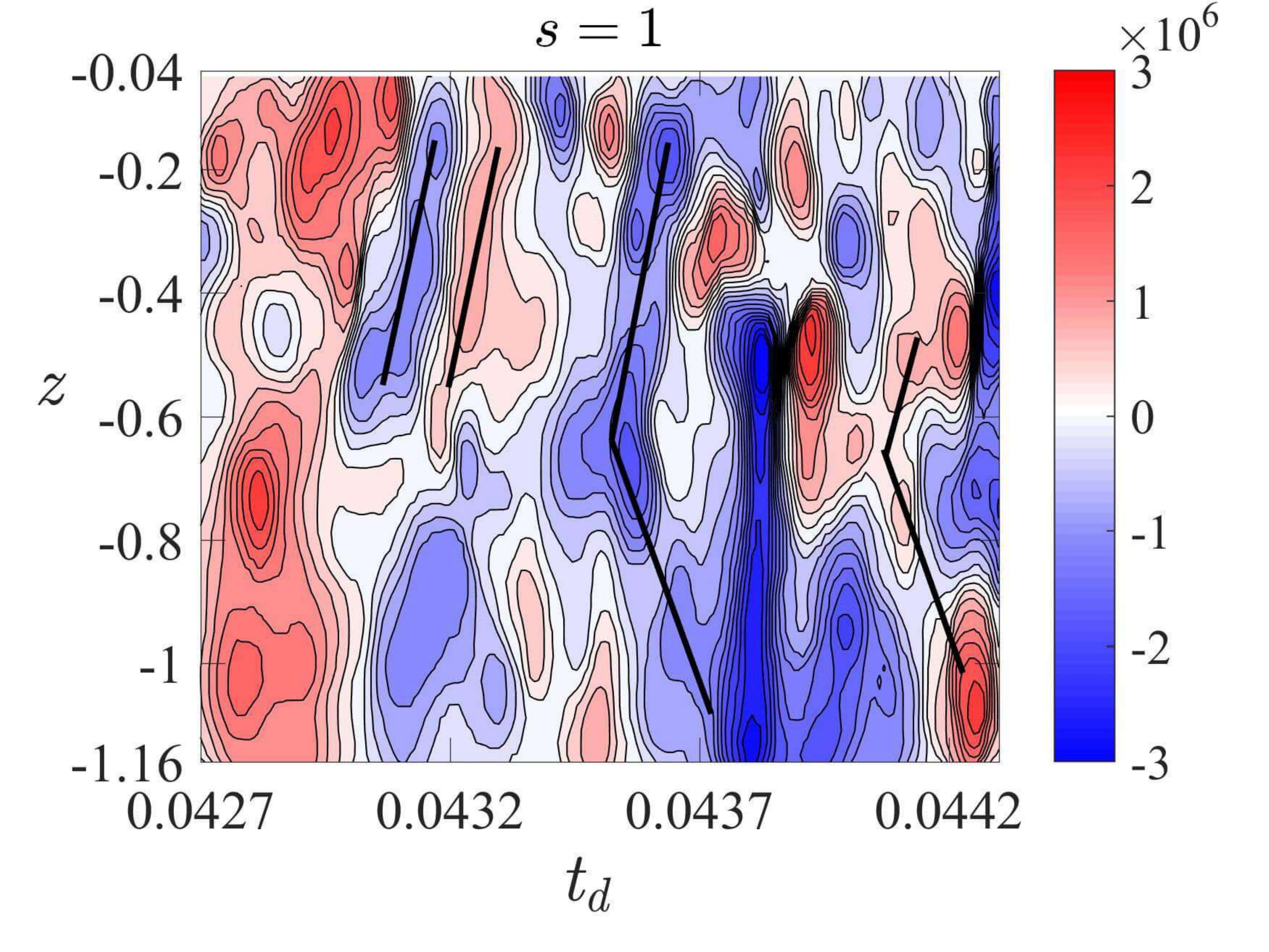}}\\
	\end{minipage}
	\hspace*{-0.3cm}
	\begin{minipage}[t]{0.5\linewidth}
		\centering
		\subfloat[	\label{fig:GV_Ra2000_b}]{\includegraphics[width=1\linewidth]
			{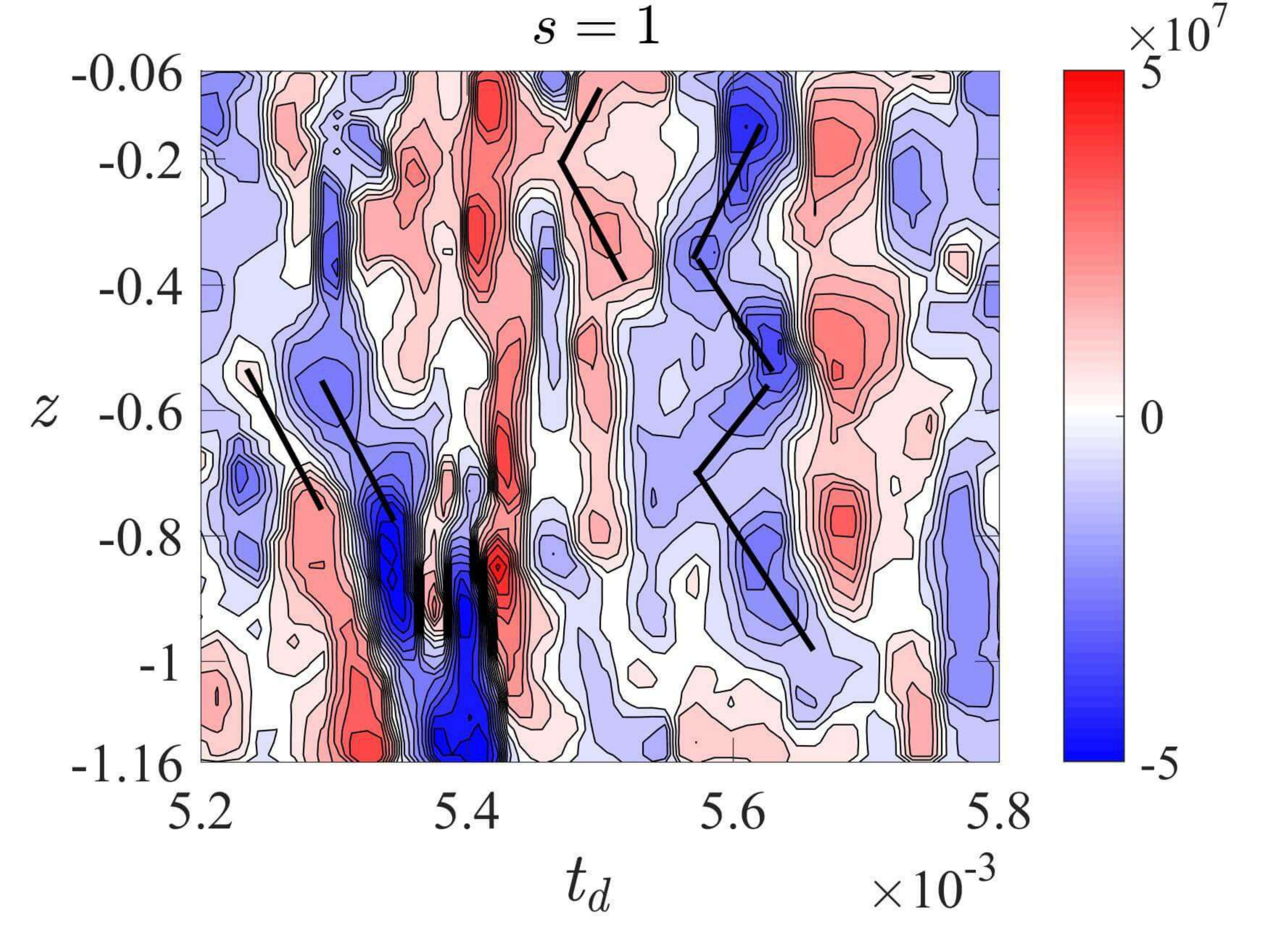}}		
	\end{minipage}
	\\	
		
	\caption{ (a) Contour plot of $\dot{u_{z}}$ at for  $l \leq 40$
	and the parameters $E=1.2 \times 10^{-5}$, $Ra=2000$,
	$Pr=Pm=5$. (b)
		$\dot{u_{z}}$ for  $l \leq 46$
	and the parameters $E=1.2 \times 10^{-5}$, $Ra=15000$,
	$Pr=Pm=5$.
	The cylindrical radius $s$ at which each
		plot is produced is given above the respective
		panel.
		The black lines indicate the direction of travel of 
		the wave and their slope gives the measured group
		velocity. The estimated
		group velocity of the fast and slow MAC waves ($U_f$ and
		$U_s$ respectively)  and the measured group velocity
		$U_{g,z}$ are given in table
		\ref{tab:gv}.}
	\label{fig:GV_Ra2000}
}
\end{figure}

\begin{figure}

{
	\centering
	
	\subfloat {\includegraphics[width=0.5\linewidth ]{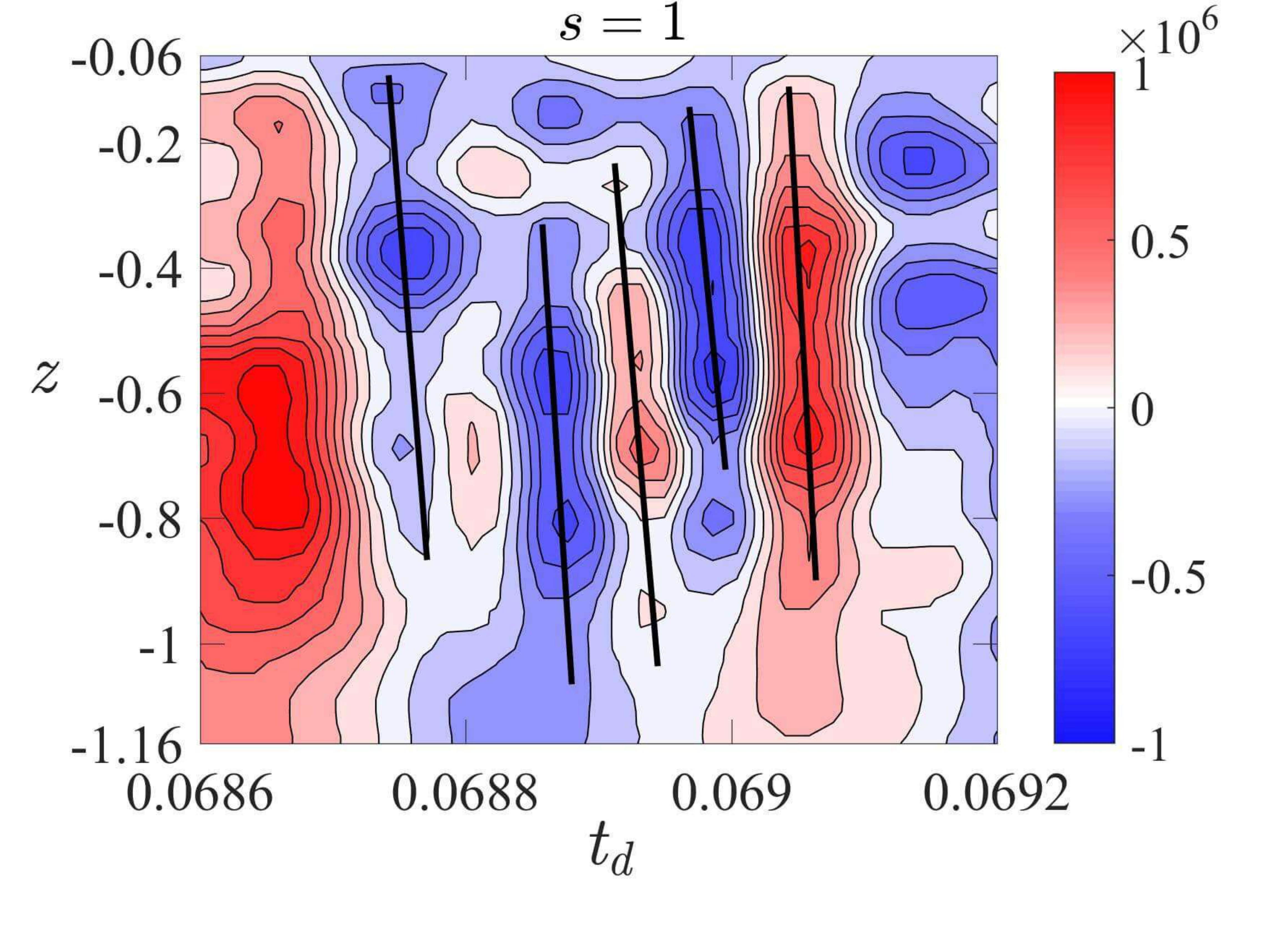}}
	
	\caption{Contour plot of $\dot{u_{z}}$ for the scales $l \leq 42$
		in a kinematic dynamo simulation with the
		parameters $E=1.2 \times 10^{-6}$, $Ra=400$, $Pm=Pr=1$.
		The black
		 lines indicate the direction of travel of 
		the wave and their slope gives the measured
		group velocity. Similar plots are obtained
		for any time window in the simulation. The estimated
		group velocity of the fast and slow MAC waves ($U_f$ and
		$U_s$ respectively)  and the measured group velocity
		$U_{g,z}$ are given in table
		\ref{tab:gv}.}
	
	\label{fig:GV_kinE6}
}

\end{figure}

	
	{
%
%
%
%



%
\section{Identification of slow MAC waves in the dynamo}
 	Wave motions in the dynamo are analysed through contour
	plots of $\dot{u_{z}}$ at points on different
	cylindrical radii $s$ and 
	for small windows of time spanning the evolution
	of magnetic field in the simulation (e.g. figure
	\ref{fig:GV_t0}). Analysis of a 
	dynamo simulations in which the field increases
	from a small seed value gives a good insight
	into the conditions for the dominance of fast 
	and slow MAC waves. Attention is focused on the
	energy containing scales $l \leq l_E$, in which the
	mean azimuthal wavenumber $\bar{m}$ is calculated
	over the range where the inequality
	$|\omega_C| > |\omega_M| > |\omega_A|$ supports
	the formation of the MAC waves. The axial group velocity
	measured from the contour plots ($U_{g,z}$) 
	is compared with the estimated value ($U_f$,
	$U_s$) obtained by taking the derivative of
	the fast ($f$) and slow ($s$) wave frequencies with respect to
	$k_z$ (table \ref{tab:gv}). 
	
	Figure \ref{fig:GV_t0} shows the measurement of
	wave motion in the dynamo simulation
	with $E=1.2 \times 10^{-6}$ and $Ra=400$ at different
	time windows during the growth of the field from a 
	small seed value. The range $l \leq l_E$ narrows down
	with increasing field intensity.
	At early times where the field
	intensity is small, fast MAC waves, whose
	frequency is comparable to that of linear inertial
	waves, are dominant.
	(figure \ref{fig:GV_t0}(a) and table 
	\ref{tab:gv}). Here, the magnetic field
	is too weak to excite slow MAC waves. 
	The group velocity measurements	confirm the presence of 
	slow MAC waves in the large scales as the field intensity
	increases with time in the dynamo. Slow wave parcels
	originating from points far from the equatorial
	plane ($z=0$) are seen to propagate in opposite
	directions with nearly equal velocity
	(e.g. figure \ref{fig:GV_t0}(b)). 
	While the slow waves
	co-exist with the fast waves at intermediate times
	($t_d=0.224$), the slow waves are
	are dominant close to dipole formation
	($t_d=0,275$; figure \ref{fig:GV_t0}(c) \& (d)).
	The measured group velocity $U_{g,z}$ increases with
	field intensity (table \ref{tab:gv}), which is the hallmark of
	the slow waves whose frequency $\omega_s$ increases
	with increasing $\omega_M$. Because $U_f$ is at least
	one order of magnitude higher than $U_s$, the fair
	agreement between $U_{g,z}$ and $U_s$ cannot be
	missed. The dominance of
	the fast waves for weak fields and the slow waves
	for strong fields is supported by figure
	\ref{fig:freq_butterfly}, where the fast
	Fourier transform (FFT) of $\dot u_{z}$
	is shown. 
	In line with the group velocity measurements, 
	the flow is made up of
	waves of frequency $\omega \sim \omega_f$ for
	weak fields (figure \ref{fig:freq_butterfly}(a)), 
	whereas for the strong fields close to dipole
	formation ((figure \ref{fig:freq_butterfly}(b)),
	waves of much lower frequency $\omega\sim \omega_s$
	 are active. 
	
	The contour plots of the time variation of the magnetic field
	$\dot{B}_z$ indicate that slow MAC wave motions are 
	dominant even at early times when the field is weak
	(figure \ref{fig:GV_mag}(a)). The measured group velocity
	$U_{g,z}$ is in fair agreement with the estimated slow wave velocity
	$U_s$ while the fast wave velocity $U_f$ is $O(10^2)$ higher
	(table \ref{tab:gv}). (Here, the mean wavenumbers 
	used for the theoretical estimate are those
	of the magnetic	field.) This interesting distinction
	between the wave motions of the flow and field is
	well explained by \cite{jfm21},
	who found that the induced magnetic field
	preferentially 
	propagates as slow MAC waves for a wide range
	of  $\omega_M/\omega_C \ll 
	1$ to $\sim 1$.
	 
	The signature of the slow waves in the energy-containing scales
	is also visible in strongly driven dipolar dynamos 
	(figure \ref{fig:GV_Ra2000}).  As the intensity of the
	self-generated field increases with increased forcing, the
	range of azimuthal wavenumbers $m$ over which 
	$|\omega_C|>|\omega_M|$ narrows down considerably
	(table \ref{tab:gv}). Consequently,
	the generation of helicity due to the slow MAC waves
	is weakened, which can explain why the axial dipole field
	$B^{P}_{10}$ diminishes in strength with increased forcing
	(table \ref{Tab:param}). There is, however, a growing dominance
	of fast waves in the large scales, which does not
	contribute to dipole field generation.
	At lower Ekman number $E$, one would expect the MAC wave window
	to widen as $|\omega_C|$ increases relative to $|\omega_M|$. The
	low-$E$ regime of Earth's outer core would thus support the
	axial dipole in strongly driven convection.
	Finally, we note that only linear inertial 
	waves are produced 
	in kinematic dynamo simulations which produce
	multipolar fields
	 (figure \ref{fig:GV_kinE6} and case (ix) in
	 table \ref{tab:gv}).
	
	Buoyancy-induced inertial waves 
	have been found in
	dynamo simulations though group velocity measurements
	\citep{ranjan2018internally}. 
	The present study has shown that slow MAC wave motions are
	measurable only when large scales of $l \leq l_E$ are considered.
	Within this range, the slow waves are predominantly
	generated in
	the MAC wave window, where $|\omega_C| > |\omega_M| > |\omega_A|$.
	To identify the scales
	where fast and slow MAC waves are active and  distinguishable
	from each other, a scale-dependent analysis of the dynamo
	spectrum is essential.

\begin{table*}
\caption{Summary of the data for MAC wave identification in the
dynamo simulations. Scales given by $l \leq l_{E}$, where $l_E$ is the
mean harmonic degree of energy injection, are considered in each
case. The sampling frequency $\omega_n$ is chosen to ensure that the
fast MAC waves are not missed in the measurement of group
velocity. The values of $\omega_{M}^2$, $\omega_{A}^2$\ and
$\omega_{C}^2$ are calculated from \eqref{om2} using the mean
wavenumbers $\bar{m}$, $\bar{k}_s$ and $\bar{k}_z$.
The measured group velocity in the $z$ direction ($U_{g,z}$)
may be compared with the estimated fast ($U_f$) or slow ($U_s$)
MAC wave velocity, as appropriate. $^*$Case (ix) is a 
kinematic dynamo simulation, which does not produce an axial dipole.}	
\begin{minipage}{1\linewidth }
	\centering
	
	\resizebox{\columnwidth}{!}{
	\begin{tabular}{c p{0.8cm} p{0.8cm} p{0.8cm} p{0.8cm} c c c c p{0.8cm} p{0.8cm} p{0.8cm} c c c}
		
		S.No.& $E$ & $Ra$ & Fig. no. & $\omega_{n}$\hspace{0pt}$(\times 10^5)$& Scales & $\bar m$ &$\bar k_{s}$ & $\bar k_{z} $& $\omega_{M}^2$\hspace{0pt}$(\times 10^9)$ &$-\omega_{A}^2$\hspace{0pt}$(\times 10^9)$ & $\omega_{C}^2$\hspace{0pt}$(\times 10^{9})$& $U_{s}$ & $U_{f}$ & $U_{g,z}$ \\
\hline
		(i) & $1.2 \times 10^{-6}$ & $400$
		 & \ref{fig:GV_t0}(a)  &6.67 & $l \leq 42$  &$12$ &$ 3.14 $& $2.51$ &$0.28$&$0.31$&$27.3$ & $- $ & $63149$ & $57333$ \\	
	    (ii) &  &  & \ref{fig:GV_t0}(b)  &6.67 & $ l \leq 36 $  &$20$ &$ 3.13 $& $2.51$ &$1.1$&$0.33$&$10.5$ & $2741 $ & $37434$ & $2133$ \\		
		(iii)&  & &\ref{fig:GV_t0}(c)  & 5& $l \leq 31 $ &$10$ &$ 3.05$& $2.17$ &$2.79$&$0.31$&$28.6$ & $4054 $ & $70749$ & $4864$ \\
		(iv)&  & &  \ref{fig:GV_t0}(d)  &5 &$ l\leq 31$ &$10$ &$ 3.05 $& $2.11$ &$3.66$&$0.31$&$27.1$ & $5295 $ & $69765$ & $6534$ \\
		(v)   &  & &  \ref{fig:GV_mag}(a)   &6.67 & $l \leq 42$  &$11$ &$ 3.61 $& $1.43 $ &$3.9$&$0.30$&$10.4$ & $780 $ & $69137$ &$966$ \\
		(vi)   &  & & \ref{fig:GV_mag}(b) &5 & $ l \leq 31$   &$9$  & $3.45$ & $1.67$& $1.68$ &$0.29 $&$23.4$ & $3625 $ & $78988$ & $3750$ \\
		\hline
				(vii)& $1.2 \times 10^{-5}$ & $2000$ & \ref{fig:GV_Ra2000}(a)   &3.33 & $l \leq 40$   &$12$ &$ 4.24 $& $2.34$ &$4.25 $&$0.74$&$5.7$ & $5071 $ & $25957$ & $6100$ \\
		\hline
		(viii)& $1.2 \times 10^{-5}$ & $15000$ &  \ref{fig:GV_Ra2000}(b)     &1.67 & $l \leq 46$ &$4$ &$ 4.78 $& $2.63$ &$12.9$&$0.36$&$31.2$ & $4856$ & $43656$ & $5187$ \\
		\hline
		(ix)$^*$ &  $1.2 \times 10^{-6}$ &  $400$  & \ref{fig:GV_kinE6}    &10 & $ l \leq 42$   &$ 16 $& $4.95$ &$2.64$ & - &  - &$1.68$ &  - &$48396$ & $55000$ \\	
		\hline

	\end{tabular}
}
\end{minipage}
\label{tab:gv}
\end{table*}

%
%
%
%

\subsection{Non-axisymmetric Alfv\'en waves}
	The generation of MAC waves in the dynamo
	is accompanied by non-axisymmetric waves along
	the cylindrical radius whose group velocity
	matches with that of Alfv\'en waves. The frequencies of
	waves that propagate orthogonal to the axis of
	rotation -- obtained by letting $\omega_C=0$ in
	equations \eqref{ndr1} and \eqref{ndr2} -- 
	would be Alfv\'enic for strong-field dynamos
	where $|\omega_C| > |\omega_M| > |\omega_A|$.
	In the dynamo simulation at $E=1.2 \times 10^{-6}$
	and  $Ra=400$, coherent radial
	motion with estimated Alfv\'en velocities
	is only noted after diffusion time $t_{d} \approx 0.1$. 
	Since slow MAC waves are first excited at $t_d \approx 0.04$
	during the growth phase of the dynamo
	(figure \ref{fig:freq_hel}(b)),
	it is reasonable to suppose that the Alfv\'en waves
	exist as the degenerate form of the MAC waves.
	 In the
	contour plots of $\dot{u_{z}}$ given in figure 
	\ref{fig:alfven} (a) and (b), the wave velocity
	is the slope measured over small time
	windows. Figure \ref{fig:alfven} (c) shows the 
	variation of the wave velocity with
	cylindrical radius $s$ for the two time intervals
	in (a) and (b), with the earlier interval showing lower
	velocity. The peak wave velocities measured
	throughout the simulation show 
	a good agreement with the Alfv\'en velocities
	calculated from the $z$-averaged local value of 
	$B_{s}^2$.  The increase in the measured wave velocity
	with the increasing intensity of $B_{s}$ in time is
	evident in figure \ref{fig:alfven} (d). 
	The waves slow down at
	the outer boundaries where the field intensity is
	weak. As we see below, the non-axisymmetric waves
	explain the growth of $u_{z}$ in
	the $s$ direction, an essential process in dipole
	formation from a seed magnetic field. 

%
%

\section{Termwise contributions to the axial dipole}

	 To understand how wave motion influences the formation of
	  the axial dipole field through the magnetic induction equation,
	 we look at stretching and advection terms in this equation
	 which dominantly influence the dipole. 
	The relative positive and negative contributions to the dipole are
	 given by
	\begin{equation}
		\frac{\int_{V} |\bm{B}^P_{10}| \,\,
[\, . \,] \, dV}{\Gamma^{P}_{10}},
	\label{reldipole}
	\end{equation}
where $\Gamma^{P}_{10}$ is defined in equation \eqref{gamma10} and the quantity
within square brackets $[ \, .\,]$ would be one of terms given in table \ref{tab:dip_comp}.
         In cylindrical polar coordinates,
	the two terms which make the highest
	 positive contribution to the axial dipole
	 are $B_{s}\partial u_{z}/\partial s$ and 
	 $B_{s}\partial u_{s}/\partial s$.
	 A significant positive contribution is also noted for
	 the  term $B_{z}\partial u_{z}/\partial z$.
	 The terms $B_{s}\partial u_{s}/\partial s$ 
	 and $B_{z}\partial u_{z}/\partial z$ 
	 are related to the production of current 
	 coils in dynamo simulations 
	 \citep{Kageyama2008Aug,takahashi2012detailed}.
	 The term $B_{s}\partial u_{z}/\partial s$ 
	 represents axial field generation due to
	 shear of axial ($z$) flow in the radial ($s$)
	 direction. This process would be influential
	 during the growth phase of the nonlinear dynamo, where
	 columnar convection is excited through slow
	 MAC wave motions. 
	  In table \ref{tab:dip_comp}, the termwise contributions
	  to the dipole in nonlinear simulations 
	  are compared with those in a kinematic simulation at
	  $E=1.2 \times 10^{-5}$ and $Ra=140$, which also produces
	  an axial dipole. Kinematic simulations at higher $Ra$ do
	  not produce an axial dipole \citep{prf2018}, and hence cannot
	  be used for comparison with the nonlinear simulations.
	   Even in the absence of slow wave motion,
	   the term $B_{s}\partial u_{z}/\partial s$ contributes
	   positively to dipole growth in the kinematic
	   dynamo due to the growth of $B_s$. Surprisingly, 
	   the toroidal--poloidal field
	   conversion via 
	   the term $(B_{\phi}/s)\partial u_{s}/\partial {\phi}$
	   -- a dominant process in the kinematic simulation -- 
	   is absent in the nonlinear simulation
	   (table \ref{tab:dip_comp}).
	    In fact, $B^{s}_{10}$, the axial dipole part
	    of the radial field component, is negatively
	    correlated with
	    $(B_{\phi}/s)\partial u_{s}/\partial {\phi}$ in the nonlinear
	  simulation (figure \ref{fig:ub_term_comp}).
	  The contribution of this term to the overall poloidal
	  field is, however, positive, which suggests that the 
	  classical alpha effect \citep{moffatt1978}
	  is still influential in generating the
	  full poloidal field from the toroidal field.
	  We also note from table \ref{tab:dip_comp}}
	   that the dipole contributions of the
	   terms $B_{z}\partial u_{z}/\partial z$,
	 $B_{s}\partial u_{s}/\partial s$ and $B_{z}\partial u_{s}/\partial z$
	 are all oppositely signed in the nonlinear and kinematic
	 simulations.
	 
	  The advection terms influenced by wave motion
	  also make influential contributions to
	  the axial dipole. The terms 
	  $-u_{z}\partial B_{z}/\partial z$ and
	  $-u_{s}\partial B_{s}/\partial s$  increase
	  preferentially in the growth phase of the nonlinear
	  dynamo and are
	  dominant positive contributors to the dipole.

\begin{table*}
	\caption{Relative contribution (in per cent) to the axial dipole by the dominant stretching and
advection terms in the magnetic induction equation, calculated from the ratio \eqref{reldipole}. 
The values are evaluated up to dipole formation time and averaged 
for the energy-containing (large) scales in the nonlinear simulations. 
The entire range of scales is considered for the kinematic simulation, marked by the superscript *.}
		\centering
		\resizebox{\columnwidth}{!}
{
			\begin{tabular}{c  c  c  c  c  c  c  }
				$E$, $Ra$ &  &  &  &  &  & \\
				\hline
				$1.2 \times 10^{-6}, 400$ &  $B_{s}\partial u_{s}/\partial s$ & $B_{s}\partial u_{z}/\partial s $ & $B_{z}\partial u_{z}/\partial z $  & $B_{z}\partial u_{s}/\partial z  $ & $(B_{\phi}/s)\partial u_{z}/\partial \phi$ 
				& $(B_{\phi}/{s})\partial u_{s}/\partial \phi$   \\ 
& 57.6 & 52.17 & 31.04 & -38.75 & -44.79 & -59.4 \\				
				$1.2 \times 10^{-5}, 500$ &  $B_{s}\partial u_{s}/\partial s$  & $B_{s}\partial u_{z}/\partial s $ & $B_{z}\partial u_{z}/\partial z $  &  $B_{z}\partial u_{s}/\partial z  $ &  $(B_{\phi}/s)\partial u_{z}/\partial \phi$ & $(B_{\phi}/{s})\partial u_{s}/\partial \phi$  \\
 & 50.8 & 43.5 & 36.9 & -33.84 & -46.7 & -49.5\\				 
				$1.2 \times 10^{-5}, 2000$ &  $B_{s}\partial u_{s}/\partial s$ & $B_{s}\partial u_{z}/\partial s$ & $B_{z}\partial u_{z}/\partial z $  &  $B_{z}\partial u_{s}/\partial z $  & $(B_{\phi}/s)\partial u_{s}/\partial \phi$  & $(B_{\phi}/s)\partial u_{s}/\partial \phi $ \\ 
& 55.01 & 48.7 & 44.4 & -39.1 & -44.05 & -61.4 \\
				$^*$$1.2 \times 10^{-5}, 140$ &  $B_{s}\partial u_{z}/\partial s$  & $(B_{\phi}/{s})\partial u_{s}/\partial \phi$  &  $B_{z}\partial u_{s}/\partial z $ & $B_{z}\partial u_{z}/\partial z $ & $B_{s}\partial u_{s}/\partial s$  & $(B_{\phi}/s)\partial u_{z}/\partial \phi$   \\ 
& 131.2 & 88.61 & 42.78 & -49.94 & -88.3 & -102.1 \\				
				\hline
				$1.2 \times 10^{-6}, 400$ & $-u_{z}\partial B_{z}/\partial z$  & $-u_{s}\partial B_{s}/\partial s $ & $-u_{z}\partial B_{s}{\partial z} $  & $-u_{z}\partial B_{s}/\partial z  $ & $-(u_{\phi}/{s})\partial B_{z}/\partial \phi$   &$-(u_{\phi}/{s})\partial B_{s}/\partial \phi$  \\ 
& 44.95 & 40.13 & 27.50 & -45.48 & -49.91 & -51.2 \\				
				$1.2 \times 10^{-5}, 500$ & $-u_{z}\partial B_{z}/\partial z$  & $-u_{s}\partial B_{s}/\partial s $ & $-u_{z}\partial B_{s}/\partial z$ & $-(u_{\phi}/s)\partial B_{s}/\partial \phi$  & $-(u_{\phi}/s)\partial B_{z}/\partial \phi$  & $-u_{s}\partial B_{z}/\partial s  $    \\ 
& 52.63 & 41.98 & 30.02 & -43.10 & -48.61 & -50.09 \\
				$1.2 \times 10^{-5}, 2000$ & $-u_{z}\partial B_{z}/\partial z$  & $-u_{s}\partial B_{s}/\partial s $ & $-u_{z}\partial B_{s}/\partial z $  & $-(u_{\phi}/s)\partial B_{s}/\partial \phi$ & $-u_{s} B_{z}/\partial s  $ & 
$-(u_{\phi}/s)\partial B_{z}/\partial \phi$\\ 
& 43.9 & 31.5 & 24.5 & -34.57 & -45.61 & -47.18 \\
				$^*$$1.2 \times 10^{-5}, 140$ & $-(u_{\phi}/{s})\partial B_{z}/\partial \phi$ & $-u_{s}\partial B_{s}/\partial s  $  & $-u_{z}\partial B_{z}/\partial z$  & $-u_{z}\partial B_{s}/\partial z $ & $-u_{s}\partial B_{z}/\partial s $  & $-(u_{\phi}/s)\partial B_{s}/\partial \phi$ \\ 
& 113.25 & 87.34 & 22.51 & -40.59 & -63.1 & -118.6 \\
				\hline
			\end{tabular}
}
	\label{tab:dip_comp}
\end{table*}

%
%
%
%
\section{Concluding remarks}
\label{concl}
      The formation of the axial dipole field in a planetary
      dynamo is strongly dependent not only on the rotation of
      the planet but also the self-generated magnetic field
      within its core. As suggested by earlier studies \citep{11sreeni,prf2018},
      the role of the magnetic field in
      dipole formation is well understood from
      dynamo models that follow the evolution of the magnetic
      field from a small seed state. At early times of evolution,
      the fast MAC waves, whose frequency is close to that of
      linear inertial waves, are abundantly present. As the field
      exceeds a threshold, marked by $|\omega_M| > |\omega_A|$,
      slow MAC waves appear; however, it is only when the field
      is strong enough to have $|\omega_M/\omega_C| \sim $ 0.1
      that the slow waves have a dominant presence in the
      dynamo (table \ref{lecomp} and figure \ref{fig:GV_t0}(c)).
      The value of $|\omega_M|$
      here must be based on the peak rather than the
      root mean square value of the field, for
      the so-called MAC wave window that satisfies the
      inequality $|\omega_C| > |\omega_M| > |\omega_A|$
      does not otherwise exist in the energy-containing scales
      of the dynamo. 
      A recent study on the evolution of isolated
      blobs subject to this inequality \citep{jfm21} 
      indicates that the
      local Elsasser number,
      \begin{displaymath}
      \varLambda \sim \left(\dfrac{\omega_M^2}{\omega_C \omega_\eta}\right)_0,
      \end{displaymath}
      would likely be $O(10^2)$ 
      for parity  between the intensities of fast and slow wave
      motions. The subscript '0' here refers to the \lq\lq isotropic"
      state of the blob that is released into the flow by buoyancy.
      In other words, the leading-order slow MAC wave frequency
      $\omega_s$ would be $O(10^2)$ times the magnetic diffusion
      frequency $\omega_\eta$.
      The peak value of $\varLambda$ in simulations at
      $E \sim 10^{-6}$ vary from $O(10^1)$--$O(10^2)$ as the 
      dynamo field increases towards the saturated state
      (figure \ref{fig:els_peak}). The instantaneous value
      of $\omega_s/\omega_\eta$ is higher than $\varLambda$ due
      to the anisotropy of the convection as 
      blobs elongate to form columns aligned with the axis of
      rotation. We anticipate that simulations at lower $E$ would 
      give $\varLambda$ of $O(10^2)$ for a wider
      range of $|\omega_M/\omega_C| \sim$ 0.1 than
      in this study. The large local value of $\varLambda$
      supports the localized excitation of slow magnetostrophic
      waves at several points in the large scales of
      spherical harmonic degree
      $l \leq l_E$, even as a global geostrophic balance exists
      at these scales \citep[e.g.][]{aurnou2017}. The generation
      of dynamo helicity -- of the same order of magnitude as the
      nonmagnetic helicity (figure 
      \ref{fig:hel_curvefit}(a) and (b)) -- is consistent
      with the excitation of the slow waves at these scales.

	  An interesting aspect of dipole field generation through
	  wave motion is that of poloidal--poloidal field conversion
	  via the term $B_{s}\partial u_{z}/\partial s$ in the
	  induction equation. While this term contributes to
	  dipole formation at low $Ra$ in kinematic dynamos through
	  the monotonic increase of $B_s$, its effect is more
	  pronounced in the nonlinear dynamo over a wide range
	  of $Ra$, where the generation
	  of radial gradients of $u_z$ happens through the
	  radial propagation of columnar vortices at the Alfv\'en
	  speed.  The twisting of the toroidal field by the radial
	  motion makes a strongly positive contribution to the
	  poloidal dipole field
	  in the kinematic dynamo, whereas it extracts energy from
	  the dipole field in the nonlinear dynamo
	  (figure \ref{fig:ub_term_comp}).

	  Since the present study has largely focused on the
	  formation of the axial dipole through magnetostrophic 
	  waves, moderately driven dynamos where
	  $|\omega_A| < |\omega_M|$ have been analysed in
	  detail. This regime is motivated in part by the 
	  thermally convecting core of early Earth, which would
	  have produced an axial dipole from a chaotic multipolar
	  field \citep{prf2018}. The stronger
	  self-generated field that accompanies stronger forcing
	  in numerical dynamos 
	  narrows down the MAC wave window in the large scales,
	  although this would not shut down the MAC waves in the
	  rapidly rotating, low-$E$ core. If forcing
	  is so strong that $|\omega_A| \sim |\omega_M|$, then
	  the slow MAC wave frequency would be considerably
	  attenuated. Consequently,
	  the helicity associated with the slow waves would
	  diminish relative to that of the fast waves, which are
	  practically unaffected by the strength of forcing. If
	  geomagnetic reversals are indeed buoyancy-driven
	  \citep{14sreeni}, then the attenuation of the slow waves
	  should provide a useful constraint on the parameter space
	  that admits reversals.
\section*{Acknowledgments}
This study was supported by Research Grant MoE-STARS/STARS-1/504 under Scheme for
Transformational and Advanced Research in Sciences awarded by the Ministry of Education, India. The computations were performed on SahasraT, the Cray XC-40
supercomputer at IISc Bangalore.

\bibliographystyle{elsarticle-harv} 
\bibliography{mac1}
\newpage


\begin{figure}
{
	\centering
	\subfloat[	\label{fig:alfven_a}]{\includegraphics[width=0.5\linewidth]{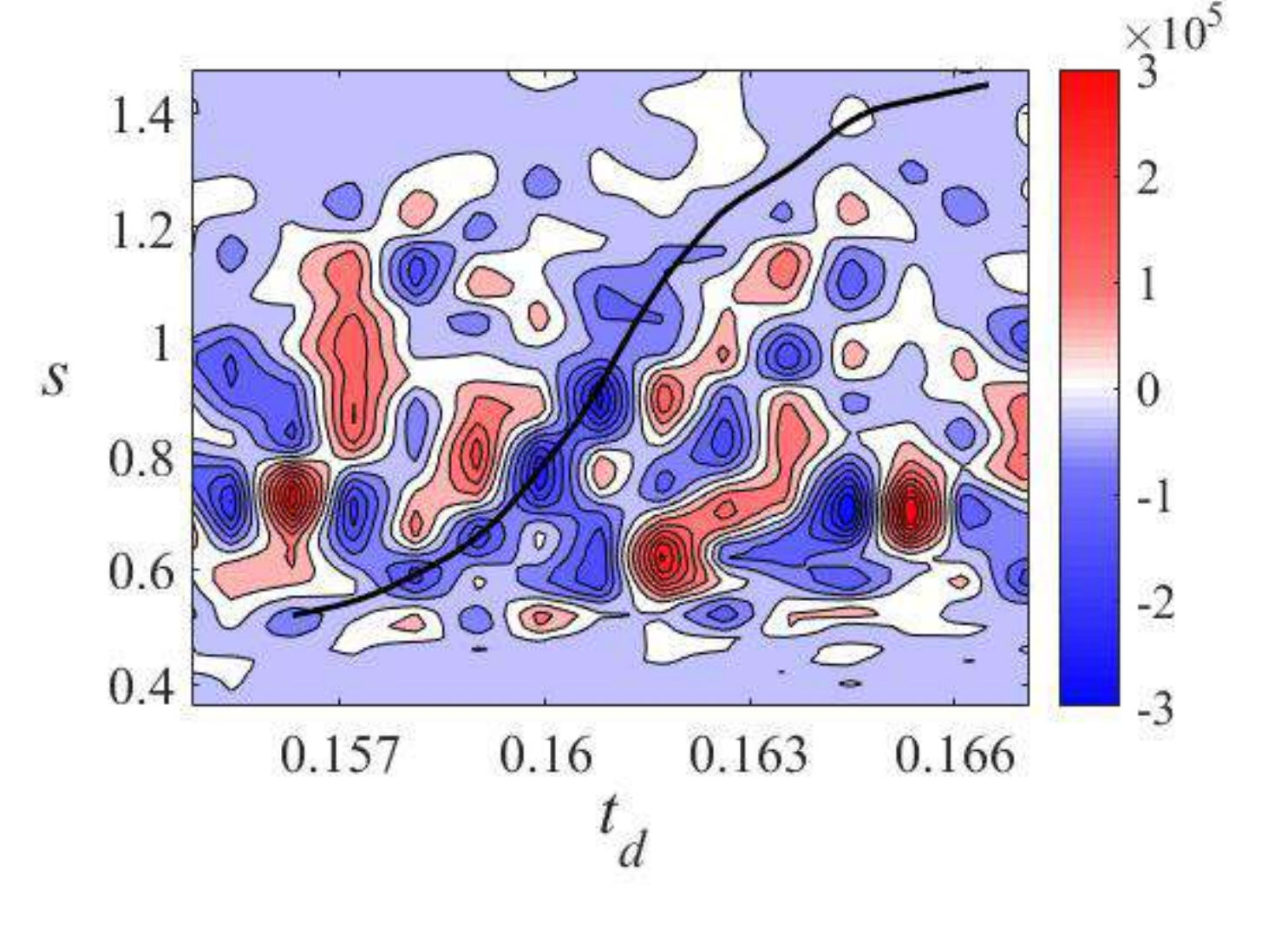}} 
	\subfloat[	\label{fig:alfven_b}]{\includegraphics[width=0.5\linewidth]{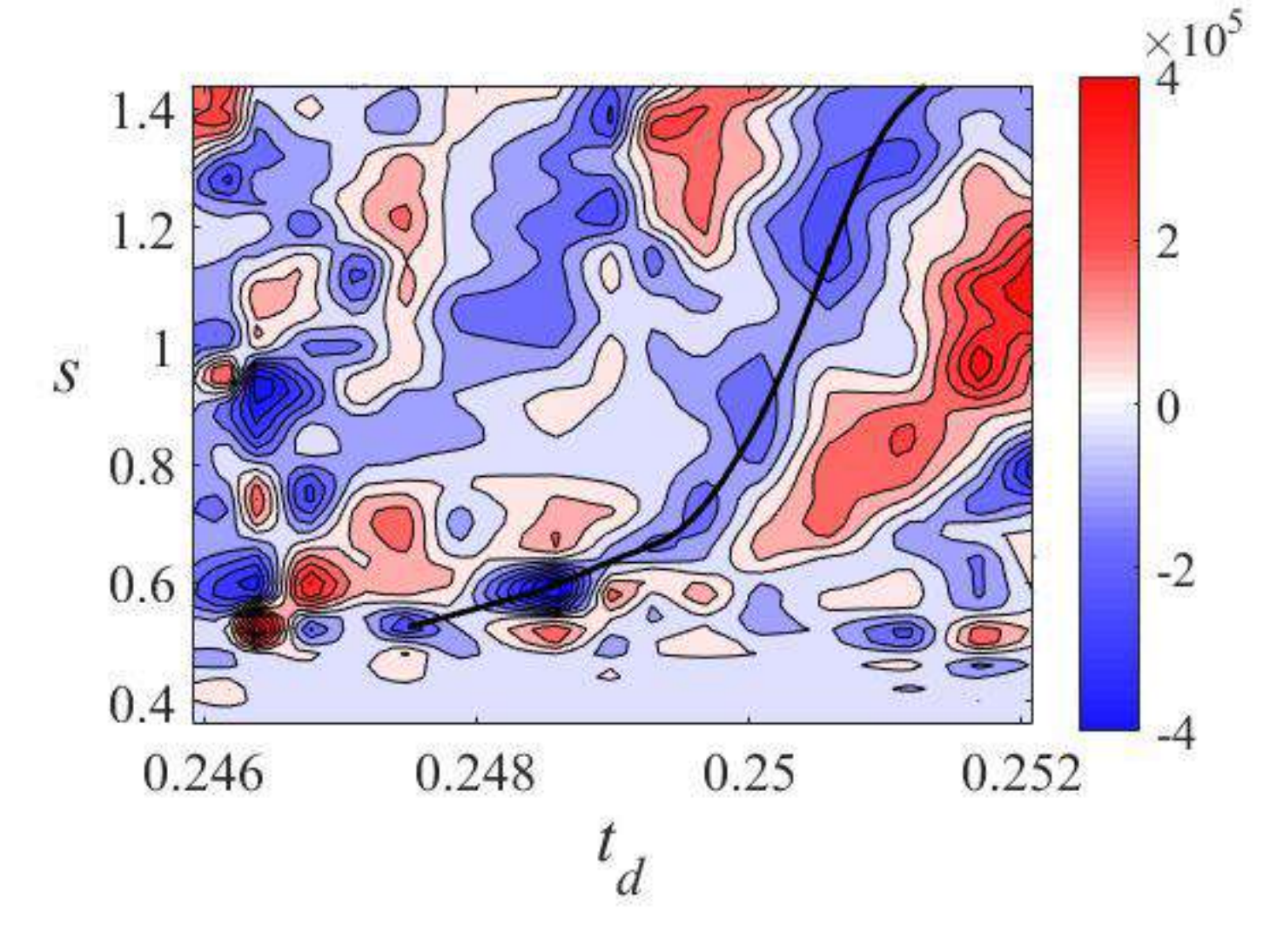}}\\
	\vspace{-1cm}
	\subfloat[	\label{fig:alfven_c}]{\includegraphics[width=0.5\linewidth]{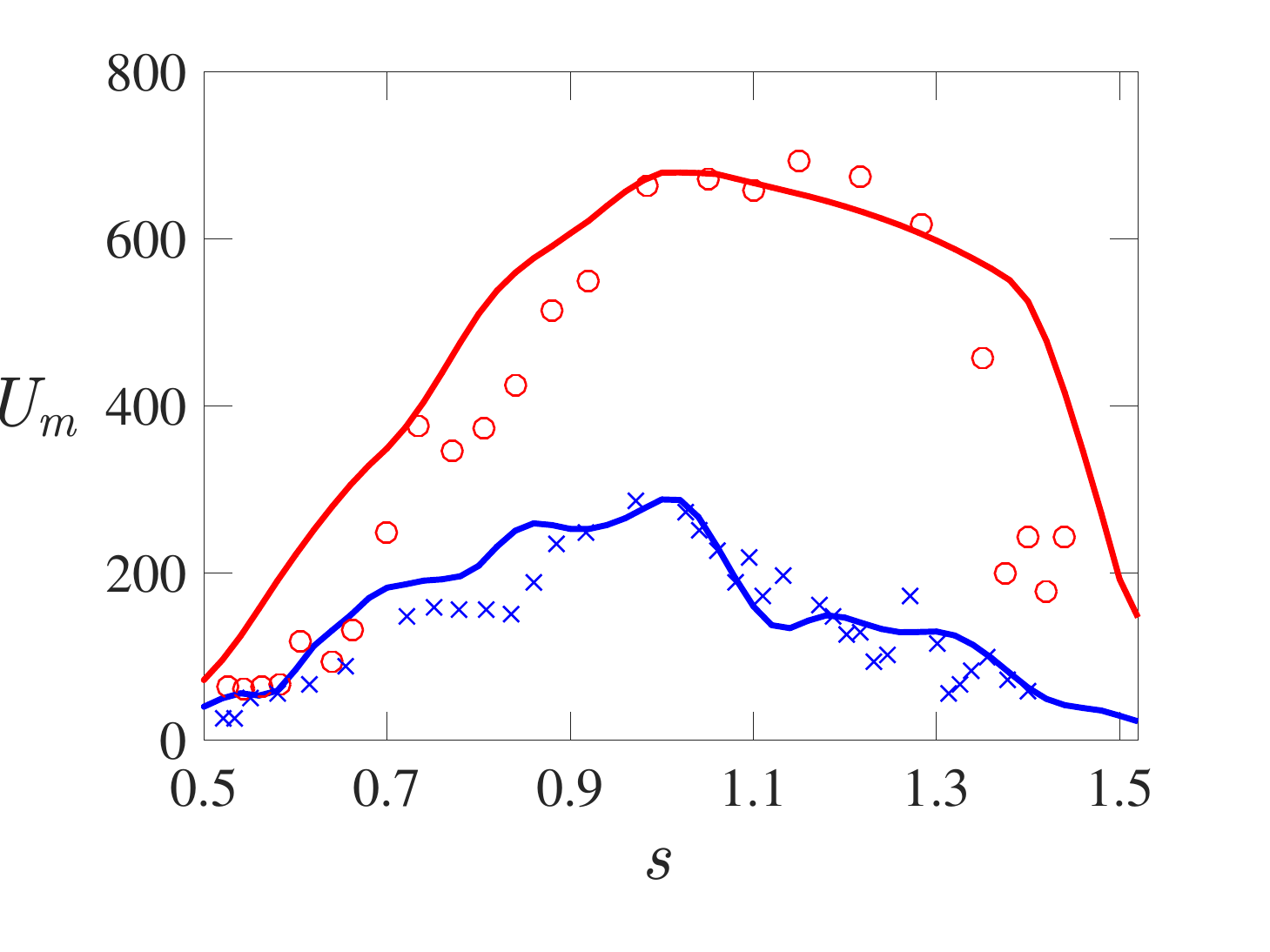}}
	\subfloat[	\label{fig:alfven_d}]{\includegraphics[width=0.5\linewidth]{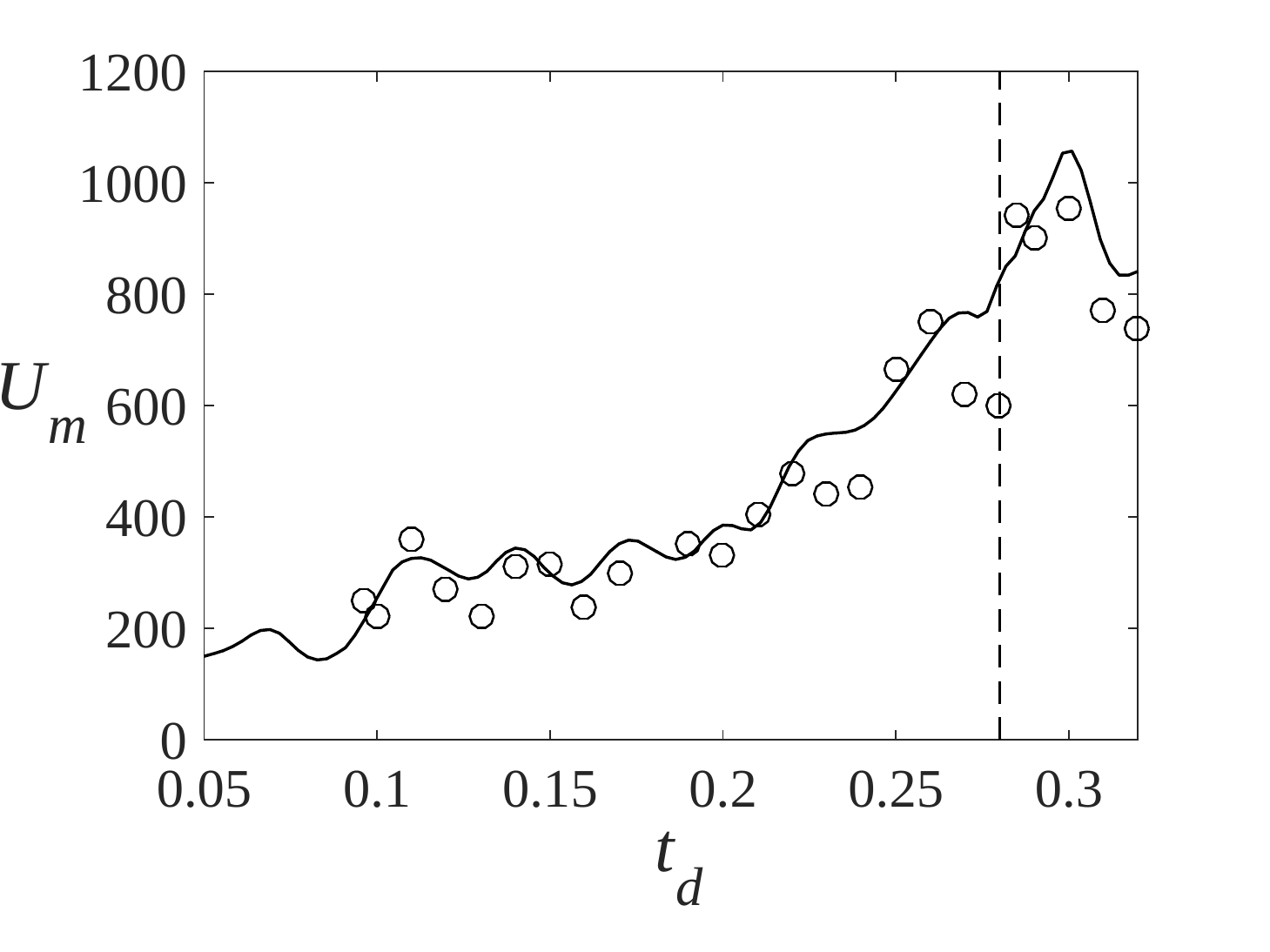}}

	\caption{(a) and (b) Contour plots of $\dot u_{z}$ for
		two time intervals for the large scales of
		$l \leq 31$. The group velocity of the waves
		is measured from 
		the slope of the black line. (c) Comparison of 
		estimated (theoretical)
		 and measured velocities at each instant 
		of time for the time intervals in (a) and
		(b), shown in blue and red respectively.
		The solid line gives the estimated
		velocity and symbols represent the measured
		values. (d) Comparison of peak velocities measured
		in the simulations at various points in the equatorial
		cross-sections. The symbols show the peak measured
		velocity of $\dot u_{z}$ while the black line shows
		the estimated velocity. The black dashed line shows
		the dipole formation time. Parameters are $Ra=400$,
		$Pm=Pr=1$, $E=1.2 \times 10^{-6}$. The group velocity 
		is estimated using the local
		value of $B_{s}$ averaged over $z$. }
	\label{fig:alfven}
}
\end{figure}
\begin{figure}
		\centering	
		{\includegraphics[width=0.5\linewidth]{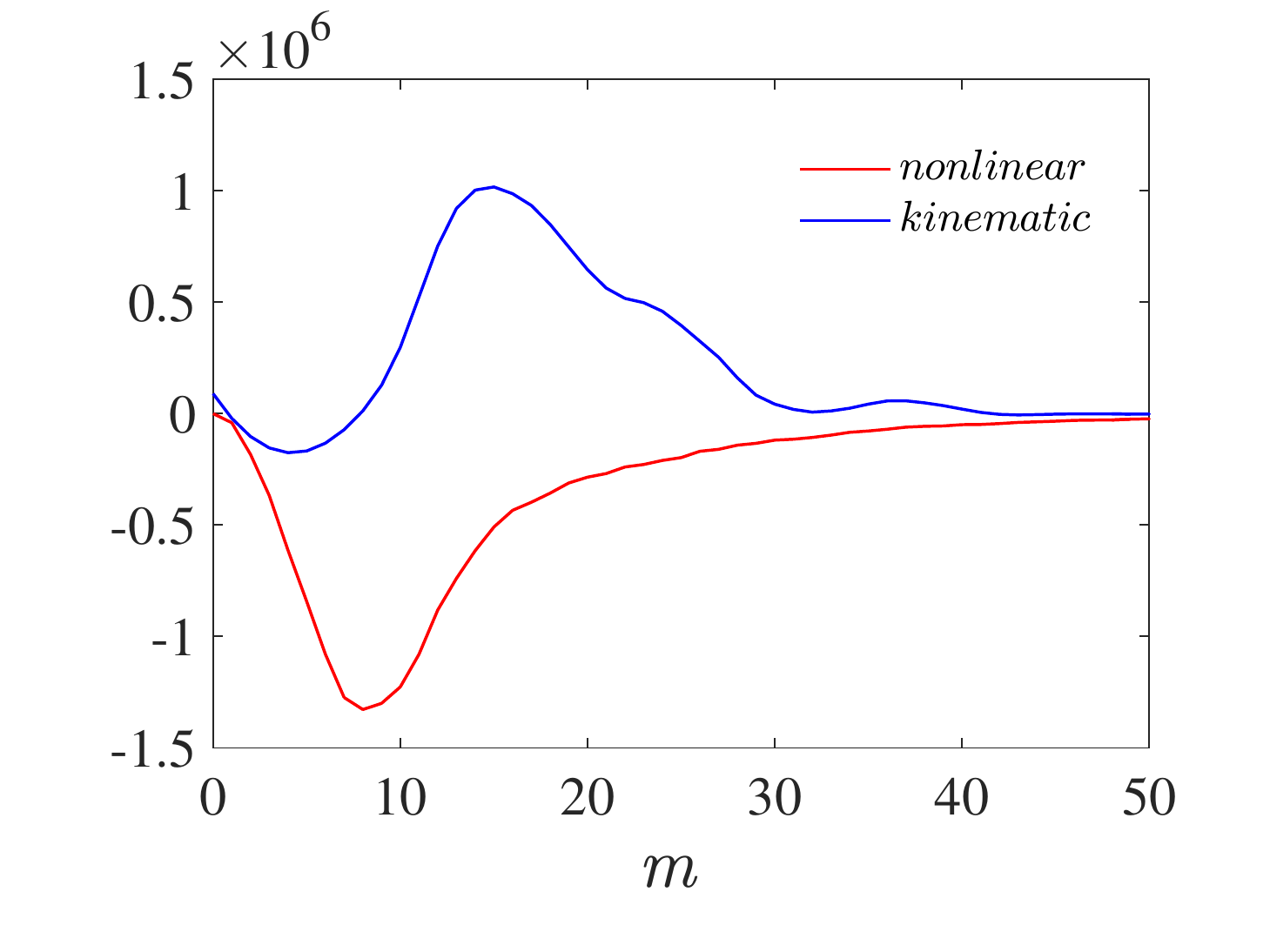}}		
		\caption{Spectral distribution of
		the contribution to the axial dipole
		energy from the term
		 $(B_{\phi}/s)\partial u_{s}/\partial \phi$
		 for nonlinear and kinematic simulations at
		 $E=1.2 \times 10^{-5}$ and $Pm=Pr=5$.
		 The nonlinear result is obtained from the saturated state at
		 $Ra=220$ whereas
		 the kinematic result is from a snapshot at $Ra=140$. }
		 \label{fig:ub_term_comp}
\end{figure}

\end{document}